\documentclass[oneside,english]{amsart}
\usepackage[T1]{fontenc}
\usepackage[latin1]{inputenc}
\usepackage{float}
\usepackage{graphicx}
\usepackage{amssymb}

 \numberwithin{equation}{section} 
 \numberwithin{figure}{section} 

\usepackage{lscape}
\usepackage[square]{natbib}
\usepackage{babel}
\makeatother
\begin{document}

\title{Permeability up-scaling using Haar Wavelets}

\author{V. Pancaldi,$^2$ K. Christensen,$^2$ P.R.
King$^1$ }

\maketitle \noindent $^{1}$Department of Earth Science and
Engineering, $^{2}$Department of Physics, Imperial College London,
United Kingdom

\begin{abstract}
In the context of flow in porous media, up-scaling is the coarsening
of a geological model and it is at the core of water resources research
and reservoir simulation. An ideal up-scaling procedure preserves
heterogeneities at different length-scales but reduces the computational
costs required by dynamic simulations. A number of up-scaling procedures
have been proposed. We present a block renormalization algorithm using
Haar wavelets which provide a representation of data based on averages
and fluctuations.

In this work, absolute permeability will be discussed for single-phase
incompressible creeping flow in the Darcy regime, leading to a finite
difference diffusion type equation for pressure. By transforming the
terms in the flow equation, given by Darcy's law, and assuming that
the change in scale does not imply a change in governing physical
principles, a new equation is obtained, identical in form to the original.
Haar wavelets allow us to relate the pressures to their averages and
apply the transformation to the entire equation, exploiting their
orthonormal property, thus providing values for the coarse permeabilities.

Focusing on the mean-field approximation leads to an up-scaling where
the solution to the coarse scale problem well approximates the averaged
fine scale pressure profile.
\end{abstract}

\section{Introduction}

The term up-scaling is used in reservoir engineering to refer to the
procedure by which a geological model is coarse grained into a flow
model. This is essential in modelling mass transport correctly to
gain an understanding of subsurface systems such as oil fields,
ground water flow and waste deposits. Correct estimation of the
transport properties of these reservoirs, including permeability, is
vital for their management. For example, in the contexts mentioned
above, good control of the fluid dynamics is necessary to ensure
optimization of recovery and the safety of the environment
\citep{Noet:futurestoch}. The procedure presented in this paper,
based on renormalization and wavelets, is a general coarse graining
technique, inspired by the wavelet treatment of the Ising model and
in line with the new developments that have been suggested in the
field of materials modelling \cite{Ism:W2005,Ism:W}.

In Section \ref{sub:Upscaling-techniques}, the main up-scaling
methods and related issues will be briefly reviewed. In Section
\ref{sec:Real-space-renormalization-and} a short account of
real-space renormalization and Haar wavelets will be given leading
to the description of the proposed method in Section
\ref{sec:A-renormalization-rule}. Numerical simulations and results
will be presented in Section \ref{sec:Numerical-simulations-and}.

The present paper is intended as a proof of concept of how wavelets
can be used in the field of upscaling by establishing a specific
formalism and applying it to the simplest cases. This allows us to
explore the underlying workings of the method, an essential step
towards the treatment of less trivial problems.

\subsection{\label{sub:Upscaling-techniques}Up-scaling techniques}

Numerous methods have been suggested for the coarsening of
permeability in geological systems, the simplest being averaging
techniques. As highlighted in a classic review \citep{Farmer:krev},
we can subdivide up-scaling methods into three categories:
deterministic, stochastic and heuristic. Further distinctions can be
made between analytical and numerical methods. The main issue with
up-scaling is the heterogeneity which characterizes natural porous
media on many different length scales. Heterogeneities range from
millimeters to kilometers, due to the great variety of types of
rocks and depositional processes that can be present in the same
system. Often, there is no clear division between the system size
and the length scales of the features or the size of the cells in
the model.

An analogy can be made between flow in porous media and currents
through resistors. This is possible because of the nature of the
equation for flow, Darcy's law, which is an elliptic equation
relating flow to the gradient of the pressure just like Ohm's law
relates current to voltage drop in conductors.

The problem of up-scaling is thus translated into solving the
Laplace-like differential equation, encouraging the application of
the wide range of methods which have been devised for this purpose
in other fields, for example field theoretical techniques,
perturbation expansions, effective medium theory, percolation
approaches or more simply finite differences and finite elements
methods, see Ref. \citep{Farmer:krev} for a recent review. A serious
drawback of these techniques, especially perturbation and effective
medium theory, is the underlying assumption that fluctuations in
permeability are small.

Renormalization offers an alternative, allowing for large
fluctuations in the system to be taken into account. Renormalization
techniques are a step-by-step approach where the system is coarsened
progressively, integrating out features on small length scales,
leading to the large scale effective permeability. Moreover,
renormalization can be applied to stochastic data sets by acting on
the probability distribution of the considered property rather than
on the single data points \cite{Hast:permup}.

With the exception of geological modelling techniques involving
object based methods and irregular grids, typically permeability
data is interpolated stochastically from the information gained at
precise locations in the reservoir. Hence, the emphasis is on
preserving the features of its statistical distribution rather than
the precise values. Furthermore, uncertainty pervades all stages of
reservoir modelling, from the measurement of permeability to the
estimation of the size of different rock type elements, rendering
statistical analysis the only viable tool to account for a range of
equiprobable scenarios which could represent the physical system
\cite{Chris:errors}.

Although there are various solutions to calculating effective
permeability for specific conditions, most of them have not been
implemented in the standard reservoir engineering packages for
industry. In practice, the methods of choice are often simple
averages, due to the ease and speed with which they can be
implemented and to the fact that precision in the estimation of
permeability in a specific location does not affect the uncertainty
implicit in the modelling process.

\section{\label{sec:Real-space-renormalization-and}Renormalization and Haar
Wavelet Transforms}

\subsection{Renormalization in up-scaling }

The concept of real-space renormalization has proved to be extremely
useful in estimating effective permeability efficiently
\citep{King:renkeff}. The basic idea behind this method is to start
with a lattice on which a property, in this case permeability, is
defined at each lattice cell. Successively the original cells are
grouped in a number of blocks, assigning new values for the
coarsened property. To avoid confusion, it is necessary to clarify
what is referred to by the words ``block'' and ``cell''. A cell is
the basic unit of the fine grid which typically characterizes the
geological model. Cell permeability is therefore what is commonly
referred to as fine permeability. A block is the basic unit of the
coarse grid used in flow simulations. The term block permeability
refers to the coarse equivalent permeability of the block,
calculated from the cell permeabilities through up-scaling
\cite{Wen:condrev}. This is clearly dependent on the boundary
conditions and is different from effective permeability, defined as
the permeability needed to relate the mathematical expectations of
the flow and of the pressure gradient. Due to the finite size of the
blocks it is only possible to consider equivalent permeability,
which ensures a match of flow patterns between the block and the
constitutive cells. After rescaling all the length scales, blocks
become cells and the result is a coarse-grained lattice with fewer
cells, but which still possesses the essential features of the
original system.

This procedure was first suggested by \cite{Kadanoff:blockspins} as
an efficient method to extrapolate the large scale behaviour of an
infinite system once fluctuations on smaller scales are averaged
out. The main advantage is that the procedure can be repeated until
the lattice has achieved the required coarseness with a low
computational cost, the algorithm being linear in the system size.

The renormalization transformation is by no means unique and many
different renormalization schemes have been proposed, some inspired
by an analogy between flow in porous media, percolation processes
and the flow of currents through resistors \cite{Will:Mathsoilren}.

Real-space transformations are a particular case of the more general
concept of the renormalization group. While the real-space version
already provides a versatile and fast technique for up-scaling, a
``full'' real- and momentum-space renormalization method for
coarse-graining of subsurface reservoirs was presented by
Hristopoulos et al. \cite{hrist:RGoverview,hrist:RG1999}. This
general treatment has confirmed the applicability of the
renormalization concept to up-scaling, providing a solution of the
problem in all orders of perturbation, even for heterogeneous
systems where large fluctuations render other methods unsound.

\subsection{The Haar wavelet transform}

The mathematical concept of wavelets was first suggested in 1909 by
Haar \cite{Haar:wav}. It found its first application in the field of
seismology in 1989 in the work of Morlet \cite{Morelt:wav}
\emph{}and has since then been at the origin of a substantial number
of new approaches to various subjects, for example, biology
\cite{Wav:bio} and statistical mechanics \cite{Ism:W}. The basic
idea underlying wavelets is to decompose a function or a set of
data, in the continuous and discrete case respectively, into basic
components and their relative coefficients \cite{Daubechies:10lec}.
\emph{}In this sense it is very similar to a Fourier transform,
where the basic components are sines and cosines and the
coefficients are given by their amplitude. Wavelet transforms,
however, offer both spatial and frequency resolution. For this
reason they have been particularly successfully applied to the
analysis of signals where it is necessary to capture both underlying
periodic functions and specific localized features, which are almost
impossible to represent with periodic components.

At this point, however, a distinction between two different uses of
wavelets must be made. On one side, wavelets can be used to compress
information in terms of reducing the number of data points with a
filtering procedure. This has been applied extensively in the
context of up-scaling by Sahimi \cite{Sahimi:swa2004}, where a
filtering process reduces the number of permeability values in the
system without compromising its general statistics. On the other
side, a more {}``pervasive'' application of wavelets can lead to the
coarsening of permeability by acting on the flow equations
themselves. This approach has been suggested in statistical physics
to compress the information relative to spins and coupling constants
in the Ising model \cite{Ism:W} and then extended to include various
aspects of materials modelling \cite{Ism:W2005}.

The main point, already noted by Best \cite{Best:LandauGins}, is
that there is a striking similarity between the perspective of
renormalization and of wavelet transforms: both highlight the
features of a system in terms of large scale behaviour and
fluctuations away from it and both provide a connection between the
different relevant scales.

In this paper, the simplest type of discrete wavelet transform, the
Haar transform, is implemented in a renormalization method. Its
effect is to separate the average of the original data from the
fluctuations, expressed in terms of differences. Wavelets are
constructed through the scaling and shifting of the so called mother
wavelet.The Haar wavelet is defined as follows,
\cite{Daubechies:10lec}:
$\psi_{jk}\left(x\right)\equiv\psi\left(2^{j/2}x-k\right)$, where
$\psi\left(x\right)$ is the mother wavelet, $j\in z$ is the scale
parameter and $k$ is the shift. This leads to a Haar wavelet matrix
of the form: \[\mathbf{H}=\left[\begin{array}{cc}1 & \phantom{-}1\\1
& -1\end{array}\right].\]

For example, if we apply this transform to a $2\times1$ vector we
can obtain a new vector in terms of sums and differences of the
original values. As will be seen in Section
\ref{sec:A-renormalization-rule} this is a useful up-scaling scheme
valid in any dimension. This is a very simple transform, however,
the formalism described can be easily applied with any matrix
transform.

\subsection{The system: single-phase laminar flow}

The simple problem analysed in this paper is single-phase creeping
flow of a viscosity dominated incompressible fluid through a porous
medium. We will assume unit viscosity and ignore the effect of
gravity. The basic equation is Darcy's equation for flow,
$\mathbf{q=-K\nabla P}$, where $\mathbf{K}$ is permeability and
$\mathbf{\nabla P}$ is the gradient of pressure, combined with the
continuity equation, $\mathbf{\nabla\cdot q}=0$, which give rise to
a Laplace-like differential equation:
$\mathbf{\nabla}\cdot(\mathbf{K\nabla P})=0$.

The discretization was performed by specifying the permeability
values at the cell centres and assuming pressure to be piece-wise
linear across the cell. Transmissibility is equal to permeability in
the case of unit volume of the discretization grid cell:
$t_{i}=k_{i}/\Delta x$, where $\Delta x=1$ is the size of the grid
cell. Assuming transmissibility $t_{i}$ to be piecewise constant
with an interface between $t_{i}$ and $t_{i+1}$ at the cell boundary
and imposing flow conservation, the inter cell transmissibility,
$t_{ij}$ is found to be the harmonic mean of $t_{i}$ and $t_{j}$
\cite{Aziz:1979}.

\begin{equation}
t_{ij}=\frac{t_{i}t_{j}}{t_{i}+t_{j}}=\frac{1}{1/t_{i}+1/t_{j}};\,\,\,\,\,\,
t_{ij}\left(t_{j}=0\right)=0;\,\,\,\,\,\,
t_{ij}\left(t_{j}=\infty\right)=t_{i}=k_{i}\label{eq:trans}\end{equation}
 As described in \cite{Aziz:1979}, this constitutes a satisfactory
approximation if the properties do not change excessively between
adjacent cells. \footnote{In the literature, the term ``block'' is
used to refer to what we call cells. Our choice is motivated to
avoid confusion given our precise definition of a block. }

Assuming permeability to be a diagonal tensor, as in an isotropic
medium, mass balance equations for the system give rise to a
five-point scheme finite-difference equation expressed in matrix
form:

\begin{equation}
\mathbf{TP=R}.\label{Eq:main}
\end{equation}
 Here, for a one-dimensional system of linear size $N$, $\mathbf{T}$
is an $N\times N$ matrix of transmissibilities, $\mathbf{P}$ is an
$N\times1$ pressure vector and $\mathbf{R}$ is an $N\times1$
boundary condition vector \cite{King:Upk,Aziz:1979}.

No-flow boundary conditions were imposed at the top and bottom of
the entire system by setting the cell permeabilities to zero, such
that also the transmissibilities in this region would be zero. A
pressure gradient in the horizontal direction was established by
setting permeability at the left and right boundaries equal to
infinity so as to generate transmissibilities at these interfaces
which are identical to the local permeabilities, see Equation
(\ref{eq:trans}). These global boundary conditions correspond to
imposing no flow at the top and bottom of the block and to a
constant pressure profile along the left and right boundaries.
Clearly, these boundaries can be rotated to calculate vertical
permeability. As outlined in \citep{Durlofsky:krev2005}, a different
choice of boundary conditions, for instance periodic, would not
alter the result significantly, given the local nature of the
up-scaling process.

\begin{figure}
 $\begin{array}{ccc}
\resizebox{0.30\textwidth}{!}{\includegraphics{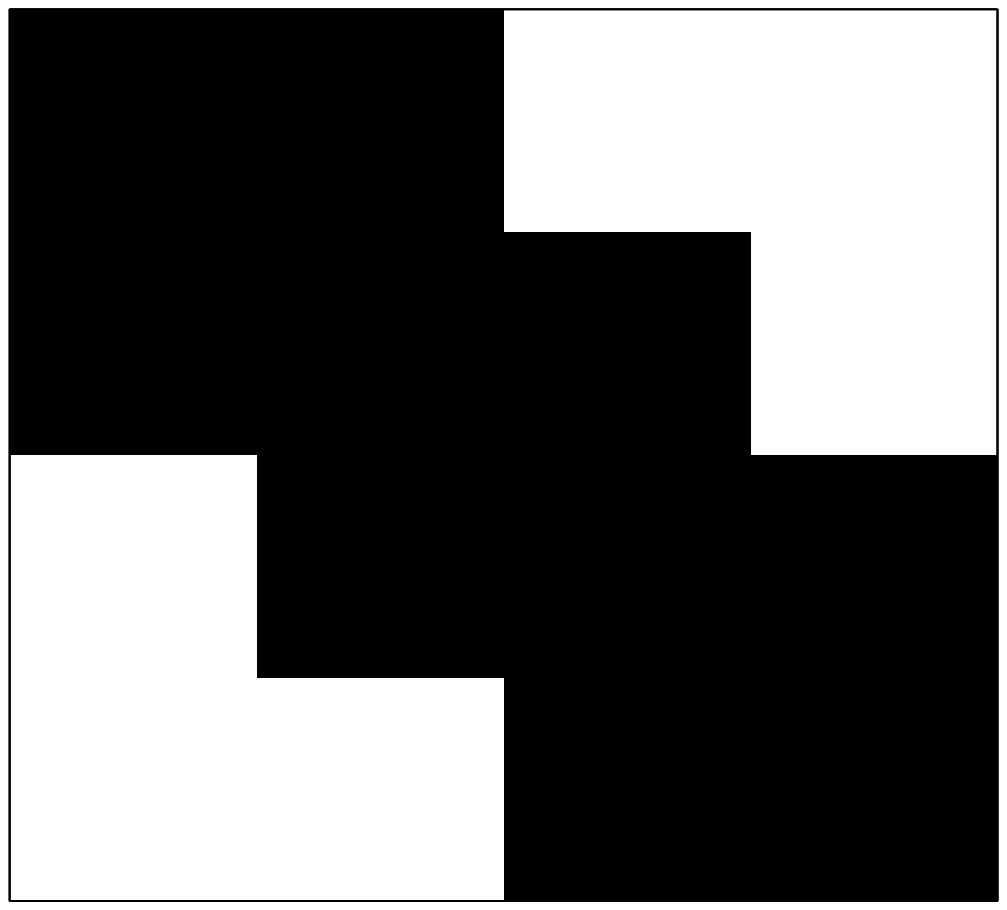}}&
\resizebox{0.30\textwidth}{!}{\includegraphics{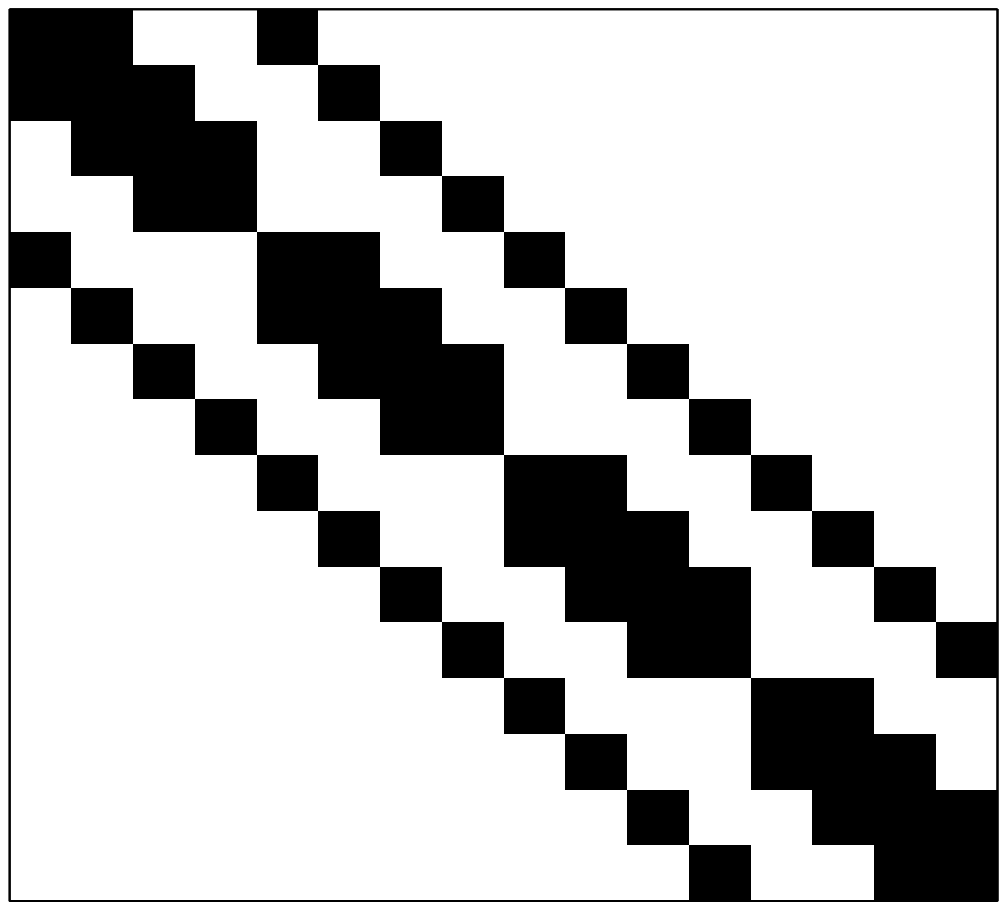}}&
\resizebox{0.30\textwidth}{!}{\includegraphics{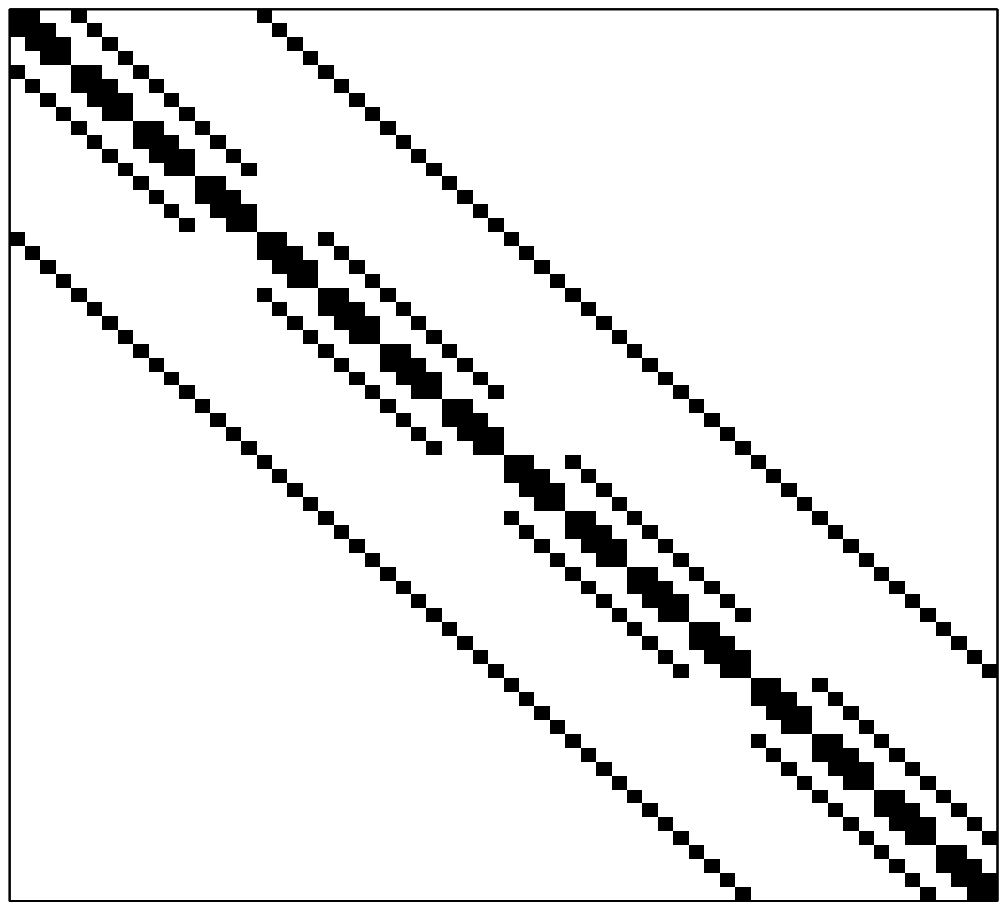}}\\
[-0.3cm] \mbox{\small (a)} & \mbox{\small (b)} & \mbox{\small (c)}
\end{array}$
\caption{ \label{cap:Transmissibility-matrix-structure}Structure of
the transmissibility matrix $\mathbf{T}$ for $N=4$ in (a) $d=1$, (b)
$d=2$, and (c) $d=3$.}
\end{figure}

In a system of dimension $d=1$, the matrix $\mathbf{T}$ has a
tridiagonal shape, arising from the coupling of each cell with its
two nearest neighbours and with itself, while in $d>1$ dimensions
further couplings are introduced leading to a diagonally dominant
sparse matrix with $2d$ non-zero off-diagonals, see Figure
\ref{cap:Transmissibility-matrix-structure}.

\section{\label{sec:A-renormalization-rule} Renormalization based on Haar
Wavelet Transforms }

\subsection{One-dimensional system}

As mentioned in Section \ref{sec:Real-space-renormalization-and},
wavelets can be used to decompose the behaviour of a system into
averages and fluctuations. For example, if we consider a
one-dimensional system consisting of two grid cells where pressure
is defined, the pressure can be expressed with the two cell values
or in terms of the average and semi-difference:

\begin{equation}
\mathbf{P'}=\mathbf{WP}=\left[\begin{array}{c}
\mathbf{\Sigma}\\
\Delta\end{array}\right],\end{equation} where the matrix
$\mathbf{W}$, the pressure vector $\mathbf{P}$, the average
$\Sigma$, and the semi-difference $\Delta$ are given by:

\begin{equation}
\mathbf{W}=\frac{1}{2}\left[\begin{array}{cc}
1 & \phantom{-}1\\
1 & -1\end{array}\right];\,\mathbf{P}=\left[\begin{array}{c}
p_{1}\\
p_{2}\end{array}\right];\,\Sigma=\frac{\begin{array}{c}
p_{1}+p_{2}\end{array}}{2};\,\Delta=\frac{\begin{array}{c}
p_{1}-p_{2}\end{array}}{2}.\label{Eq:1d}\end{equation}

The matrix $\mathbf{W}$ relates the original pressure variable
$\mathbf{P}$ to the new pressure variable $\mathbf{P'}$. Thus if we
operate on the pressure vector of Equation (\ref{Eq:main}) with
$\mathbf{W}$, a new pressure vector $\mathbf{P'}$ can be obtained,
where the first element is the average of the original pressures,
see Equation (\ref{Eq:1d}). This matrix is simply $1/2\,\mathbf{H}$,
where $\mathbf{H}$ is the Haar transform matrix for a $1\times2$
system.

Let us consider a $1\times N$ system, with $N=4$, that we want to
coarsen by a factor $n=2$ by transforming a $1\times4$ group of
cells into a $1\times2$ group of blocks. We will have:

\begin{equation}
\mathbf{W}=\frac{1}{2}\left[\begin{array}{cccc}
1 & \phantom{-}1 & 0 & \phantom{-}0\\
0 & \phantom{-}0 & 1 & \phantom{-}1\\
1 & -1 & 0 & \phantom{-}0\\
0 & \phantom{-}0 & 1 &
-1\end{array}\right];\,\mathbf{P}=\left[\begin{array}{c}
p_{1}\\
p_{2}\\
p_{3}\\
p_{4}\end{array}\right];\end{equation}

\begin{equation}
\mathbf{P'=\left[\mathbf{\begin{array}{c}
\mathbf{\Sigma}\\
\mathbf{\Delta}\end{array}}\right]};\mathbf{\,\Sigma}=\left[\begin{array}{c}
\frac{\begin{array}{c}
p_{1}+p_{2}\end{array}}{2}\\
\frac{\begin{array}{c}
p_{3}+p_{4}\end{array}}{2}\end{array}\right];\,\mathbf{\Delta}=\left[\begin{array}{c}
\frac{\begin{array}{c}
p_{1}-p_{2}\end{array}}{2}\\
\frac{\begin{array}{c}
p_{3}-p_{4}\end{array}}{2}\end{array}\right].\end{equation}

An important property of $\mathbf{W}$ is that the product
$\mathbf{WW^{T}}$ is the identity matrix multiplied by a factor of
$1/n$. $\mathbf{WW^{T}}$can be therefore inserted altering Equation
\ref{Eq:main} only by a factor of $n$:

\begin{equation}
\mathbf{TW^{T}WP}=\frac{1}{n}\mathbf{R}.\end{equation} To complete
the equation transformation we multiply by $\mathbf{W}$ on both
sides to obtain a new transmissibility matrix and a new boundary
condition vector applied to the transformed pressure:

\begin{equation}
\mathbf{\left(WTW^{T}\right)WP}=\frac{1}{n}\mathbf{WR}.\end{equation}
Defining the transformed variables,

\begin{equation}
\mathbf{T'=WTW^{T}};\mathbf{\,\,\,\,\, P'=WP};\mathbf{\,\,\,\,\,
R'=WR};\end{equation}

\noindent we have

\begin{equation}
\mathbf{T'P'=}\frac{1}{n}\mathbf{R'}.\label{eq:app}\end{equation} Up
to this point, the transformation has been completely reversible; in
fact, we have simply changed the variables with which we represent
the system. Now we approximate Equation (\ref{eq:app}) by ignoring
the fluctuations of the systems to preserve the large scale
behaviour. To do this, we define new variables
$\mathbf{\mathcal{P}}$ and $\mathbf{\mathcal{R}}$ composed of the
first $(N/2)$ elements of $\mathbf{P'}$ and $\mathbf{R'}$
respectively, and $\mathbf{\mathcal{T}}$ as the $(N/2)\times(N/2)$
upper left corner of $\mathbf{T'}$.

\begin{center}$\mathbf{T}=\left[\begin{array}{cccc}
2k_{1}+t_{12} & -t_{12} & 0 & 0\\
-t_{12} & t_{12}+t_{23} & -t_{23} & 0\\
0 & -t_{23} & t_{23}+t_{34} & -t_{34}\\
0 & 0 & -t_{34} & t_{34}+2k_{4}\end{array}\right];$\end{center}

\begin{center}$\mathbf{T'}=\left[\begin{array}{cccc}
2k_{1}+t_{23} & -t_{23} & 2k_{1}-t_{23} & -t_{23}\\
-t_{23} & t_{23}+2k_{4} & t_{23} & t_{23}-2k_{4}\\
2k_{1}-t_{23} & t_{23} & t_{23}+t_{34} & t_{23}\\
-t_{34} & t_{23}-2k_{4} & t_{23} &
4t_{34}+2k_{4}+t_{23}\end{array}\right];$\end{center}

\begin{equation}
\mathbf{T'}=\left[\begin{array}{cc}
A & B\\
B^{T} &
C\end{array}\right];\,\,\,\,\,\,\,\,\,\,\,\mathbf{\mathcal{T}}=A=\left[\begin{array}{cc}
2k_{1}+t_{23} & -t_{23}\\
-t_{23} & t_{23}+2k_{4}\end{array}\right].\end{equation}

To determine the coarse pressure, we invert the renormalised
transmissibility matrix $\mathcal{T}$ and multiply the resulting
pressure by $2$. This rescale is necessary to compensate for the
change from cell values to block values, which has doubled the size
of $\Delta x$.

\begin{equation}
\mathcal{TP}=\frac{1}{2}\mathcal{R};\,\,\,\,\mathcal{P}=\frac{1}{2}\mathcal{T}^{-1}\mathcal{R};\,\,\,\,\mathbf{P}_{coarse}=2\mathcal{P}.\label{eq:inversion}\end{equation}

\noindent Using $\mathbf{\mathcal{T}}$, $\mathbf{\mathcal{P}}$, and
$\mathbf{\mathcal{R}}$ corresponds to assuming that fluctuations of
pressures $\mathbf{\Delta}$, are negligible. In other words, we
represent the system in what is commonly called a mean-field
approximation where only the average behaviour of the pressure field
is considered. Hence, exploiting the orthonormal property of
$\mathbf{W}$, an expression for the coarse transmissibility can be
derived, by operating on Darcy's equation on the fine scale, leading
to a mean-field pressure solution. The general principle underlying
this method, can be applied in any dimension and to all problems
which require coarsening.

\subsection{Two- and three-dimensional systems}

In $d$-dimensions a similar treatment can be performed, where the
equivalent of a linear arrangement of $N$ cells is a $d$-hypercube
of linear size $N$ which we want to coarsen by a factor of $2$ in
each direction. In this case a convention for the ordering of the
pressures in the vector is needed. The coefficient in the
$\mathbf{W}$ matrix and the pressure rescale factor is now
$1/2^{d}$. Moreover, while it is easy to write down expressions for
the average and difference for two cell values, a complication
arises when cells are averaged in a dimension equal or higher than
two. In this case, the pressures are averaged 4 at a time and there
is no unique way to define their difference. For example, the
$\mathbf{W}$ matrix and $\mathbf{P'}$ for a $2\times2$ system can be
given by:

\begin{equation}
\mathbf{W}=\frac{1}{4}\left[\begin{array}{cccc}
1 & \phantom{-}1 & \phantom{-}1 & \phantom{-}1\\
1 & -1 & \phantom{-}1 & -1\\
1 & \phantom{-}1 & -1 & -1\\
1 & -1 & -1 &
\phantom{-}1\end{array}\right];\,\,\,\,\mathbf{P'}=\frac{1}{4}\left[\begin{array}{c}
p_{1}+p_{2}+p_{3}+p_{4}\\
p_{1}-p_{2}+p_{3}-p_{4}\\
p_{1}+p_{2}-p_{3}-p_{4}\\
p_{1}-p_{2}-p_{3}+p_{4}\end{array}\right],\end{equation} but this is
by no means the only valid choice. The constraints on $\mathbf{W}$
are that the top row should produce the pressures average, that
$\mathbf{WW^{T}}$ is proportional to the identity and that all rows
are orthonormal to the top one.

\begin{figure}[H]
\begin{center}\includegraphics[%
  scale=0.25]{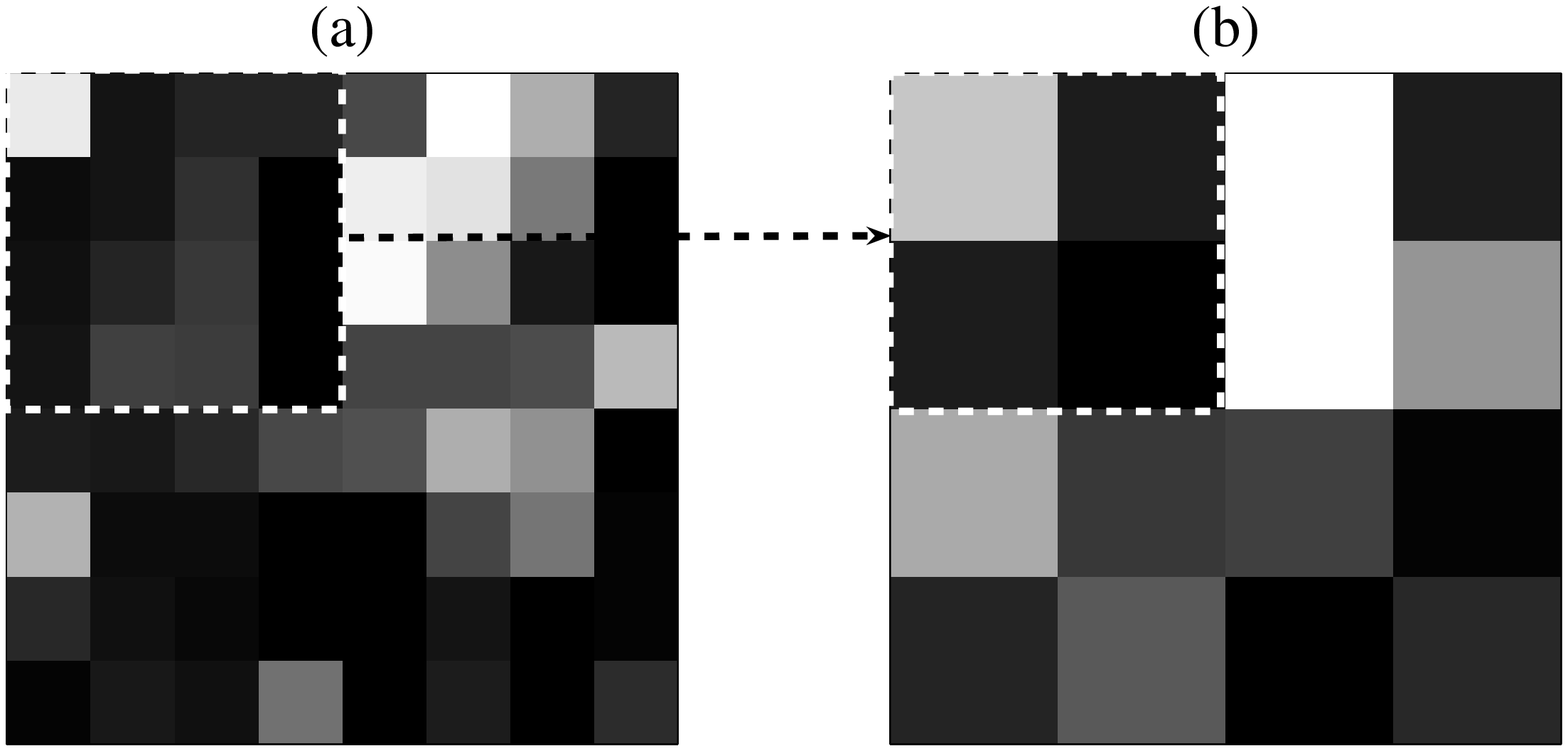}\end{center}

\begin{center}\includegraphics[%
  scale=0.4]{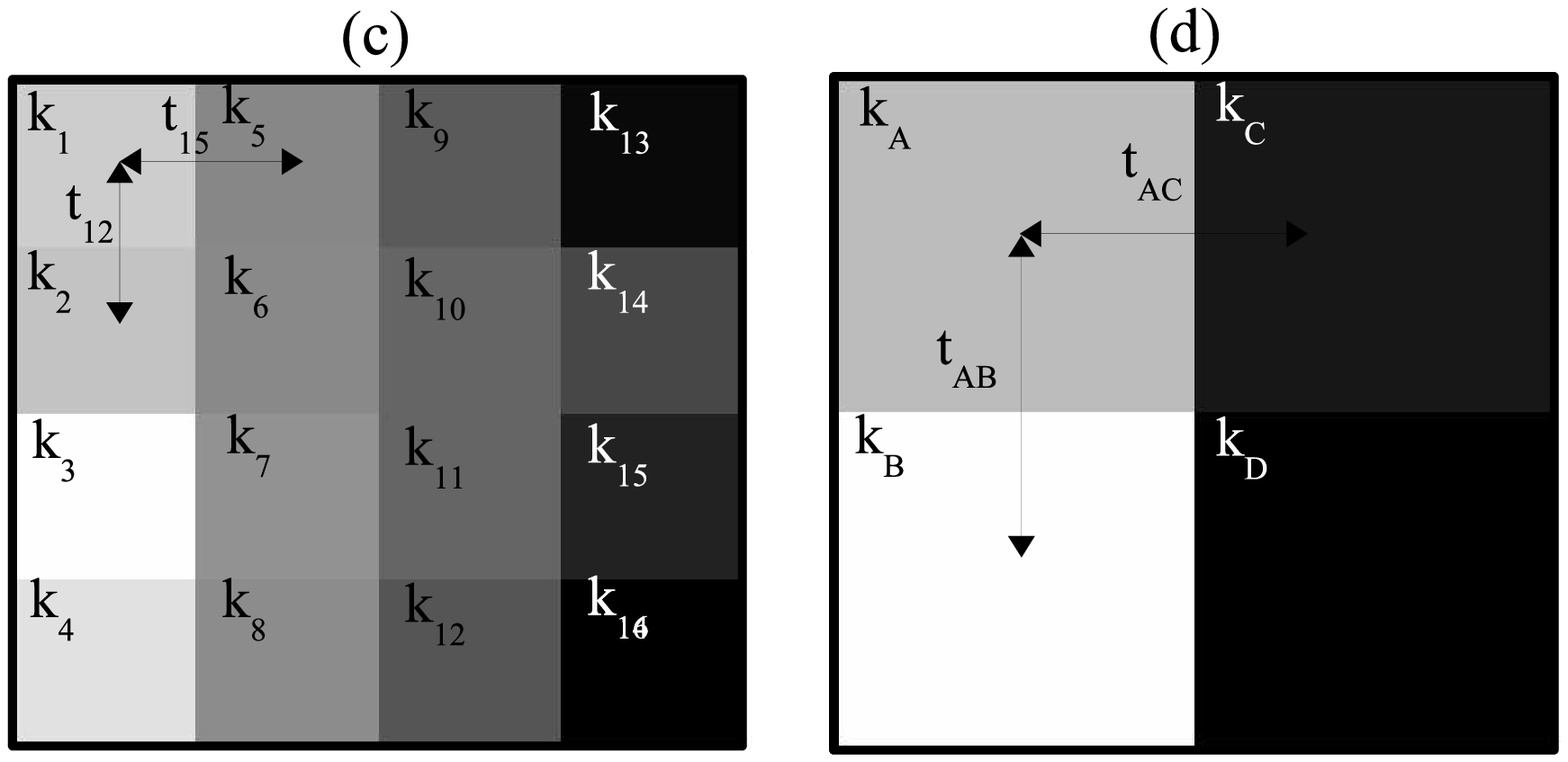}\end{center}

\caption{\label{cap:A-schematic-representation}A schematic
representation of the relation between cell and block permeabilities
and transmissibilities. One step in the renormalization algorithm.
(\textbf{a}) $8\times8$ permeability map. (\textbf{b}) The
$4\times4$ coarsened permeability map. Notice how a $4\times4$ group
of cells is substituted by a $2\times2$ group of blocks.
(\textbf{c}) Blow-up of a $2\times2$ group of cells. (\textbf{d})
Blow-up of a $2\times2$ group of blocks. \textbf{\emph{}}Properties
of cells are subscripted with numbers, properties of blocks with
letters. Permeabilities are indicated by $k$ and transmissibilities
with $t$, $k{}_{A}=\left(k_{1}+k_{2}\right)/2,\,
t{}_{AB}=\left(t_{23}+t_{67}\right)/2,\,
t{}_{AC}=\left(t_{59}+t_{610}\right)/2$.}
\end{figure}

While in one dimension the flow follows a forced path, already in
two dimensions we can recover many of the characteristics of
transport phenomena. Moreover, when looking at the elements of the
matrix $\mathbf{\mathcal{T}}$ for the two-dimensional system, it was
noted that the block permeability can be obtained by performing a
specific average of the cell permeabilities, see Figure
\ref{cap:A-schematic-representation} and Appendix.

For a $4\times4$ system, the transmissibility matrix is
$16\times16$. When transformed with $\mathbf{W}$ and
$\mathbf{W^{T}}$ the matrix obtained is still $16\times16$, but
taking the first four rows and columns only, we get a $4\times4$
matrix. This can be compared to the transmissibility matrix of a
$2\times2$ system to deduce the relation between the permeabilities
at cell and block level, see Appendix.

Accordingly, a renormalization algorithm was implemented whereby
groups of $4\times4$ cells are progressively substituted by groups
of $2\times2$ blocks, until the required degree of coarsening in
permeability is achieved. This procedure is fast and can be further
improved with the use of parallel computing. Finally,
$\mathbf{\mathcal{T}}$ is inverted to obtain the coarse pressure,
see Equation (\ref{eq:inversion}). The procedure in $d$-dimensions
is as follows:

\begin{enumerate}
\item Start with a permeability map, linear size $N$ multiple of 4. Calculate
the pressure by inverting the transmissibility matrix, see Equation
(\ref{Eq:main}).
\item Subdivide the system into groups of $4^{d}$ cells. Substitute each
group with a new group of $2^{d}$ blocks, calculating the new
permeability according to the averaging rule, see Appendix and
Figure \ref{cap:A-schematic-representation}.
\item The new system has a factor of $2^{d}$ less cells. Calculate the
upscaled pressure by inverting the new transmissibility matrix and
rescaling, Equation (\ref{eq:inversion}).
\end{enumerate}
Clearly the higher the dimension, the bigger the advantage in
avoiding a double matrix multiplication.

It should be stressed that this renormalization scheme derives
directly from the representation of the problem in the mean-field
approximation and from the choice of $\mathbf{W}$ matrix. This
result relates the elimination of permeability fluctuations to the
smoothening of fluctuations in pressure, revealing the basic
principle underlying renormalization methods for up-scaling.
Importantly, it also represents the starting point for devising a
controlled method to include the effects of fluctuations in the
coarsening process.

\section{\label{sec:Numerical-simulations-and}Numerical Simulations and Heterogeneities}

\subsection{Stochastically generated correlated permeability}

To emphasize the importance of maintaining the statistical
properties of the permeability distribution, various realizations
were generated with the same moments. Permeability was simulated as
a random, log-normally distributed correlated variable on two- and
three-dimensional Cartesian regular grids with a moving average
technique \cite{Wall:cor}.

The starting point is an uncorrelated field, that is normally or
uniformly distributed random numbers are assigned to each cell. Then
the correlation is introduced by averaging these values with a
moving circle technique \cite{Wall:cor}. By the central limit
theorem, the new distribution remains normal, at least for
sufficiently big circles, independent of the statistics of the
initial data. Moreover, the correlation length is related to the
radius of the circle used in the averaging process. Permeability is
then taken to be the exponential of this distribution. Anisotropic
systems can be generated by using ellipses to account for different
rock types in the simulated reservoir.

The upscaled pressure was compared with the simple averaging of the
fine pressure. Errors were calculated as differences between the two
pressure solutions at the same coarsening level and then averaged
over the entire system. While the average error is a useful measure
of accuracy, localizing the discrepancies allows us to look for
their justification in view of heterogeneities.

\subsection{\label{sub:Analysis-of-heterogeneity}Analysis of heterogeneity in
permeability distribution}

The simplest test cases to be analysed are two layered systems where
exact analytical solutions are known. More precisely, the equivalent
permeability for flow parallel to the strata is the arithmetic
average of the different permeabilities, and for perpendicular flow
it is the harmonic average. In general, these two averages can be
shown to be respectively the lower and upper limit on the effective
permeability of any system \cite{Farmer:krev}. As can be expected
the new renormalization technique is just as good in these cases as
others of its kind. It must be noted that while renormalization
according to the resistor analogy produces a final number
corresponding to  the equivalent permeability, the last step of the
wavelet method can only lead to a $2\times2$ cell. A number can be
obtained afterwards, but this is necessarily going to be some kind
of average. For example, in the case of vertical layering, while at
the third coarsening step the renormalization method already has a
homogeneous character, the wavelet method still has a layered
structure. The correct result, that is, the harmonic mean, can be
recovered by taking the harmonic mean of the two layers. In the case
of a chess-board configuration, where resistor analogy
renormalization underestimates permeability with an error increasing
with the difference between the two layer permeabilities
\cite{Zimm:ren_comp}, the wavelet method overestimates it by an even
larger amount. In this case it is possible to show analytically that
the exact result should be the geometric mean \cite{Farmer:krev}
while the wavelet method result is the arithmetic mean, as is
expected given the averaging which takes place in the algorithm.
This is not ideal but at least the error can be predicted and its
source clearly identified, see Table
\ref{cap:Comparison-of-wavelet}.
\begin{table}[H]
\begin{center}\begin{tabular}{llll}
\hline $k_{1}$= 2500, $k_{2}$= 5000& $k_{\textrm{resistor}}$&
$k_{\textrm{wavelet}}$& $k_{\textrm{exact}}$\tabularnewline \hline
Perpendicular layering& 3333& 3333& 3333\tabularnewline Parallel
layering& 3750& 3750& 3750\tabularnewline Chess-board& 3429& 3750&
3535.5\tabularnewline \hline
\end{tabular}\end{center}

\caption{\label{cap:Comparison-of-wavelet}Comparison of effective
permeability obtained by resistor and wavelet based renormalization
for layered and chess-board systems with cells of low ($k_{1}$) and
high ($k_{2}$) homogeneous permeability reduced to a single cell.
Both methods predict the exact results correctly for the layered
cases while both fail in the chess-board case. }
\end{table}

Initially two-dimensional systems were analysed so the method
described will refer to this case. Results are also presented in
three dimensions, where the procedure is identical in concept.

First, an analysis was made on the permeability distribution at each
up-scaling step, see Table \ref{cap:Distribution-of-permeabilities}.

\begin{table}[H]
\begin{center}\begin{tabular}{ccc}
\hline Cell size& Mean& Std\tabularnewline \hline $\phantom{1}$1&
4902.9& 11.8597\tabularnewline $\phantom{1}$2& 4902.8&
11.54\tabularnewline $\phantom{1}$4& 4902.9& 11.05\tabularnewline
$\phantom{1}$8& 4903.1& 10.24\tabularnewline 16& 4903.9&
$\phantom{1}$8.36\tabularnewline \hline
\end{tabular}\end{center}

\caption{\label{cap:Distribution-of-permeabilities}Statistics of
permeability distribution at each coarsening step, for a
$64\times64$ system, with correlation length $r=10$ averaged over 10
realizations. At each up-scaling step the cell size doubles. Notice
how the renormalization preserves the mean and how the standard
deviation starts to decrease considerably only when the cell size is
comparable to the correlation length.}
\end{table}

\begin{figure}[H]
\begin{center}\includegraphics[%
  scale=0.25]{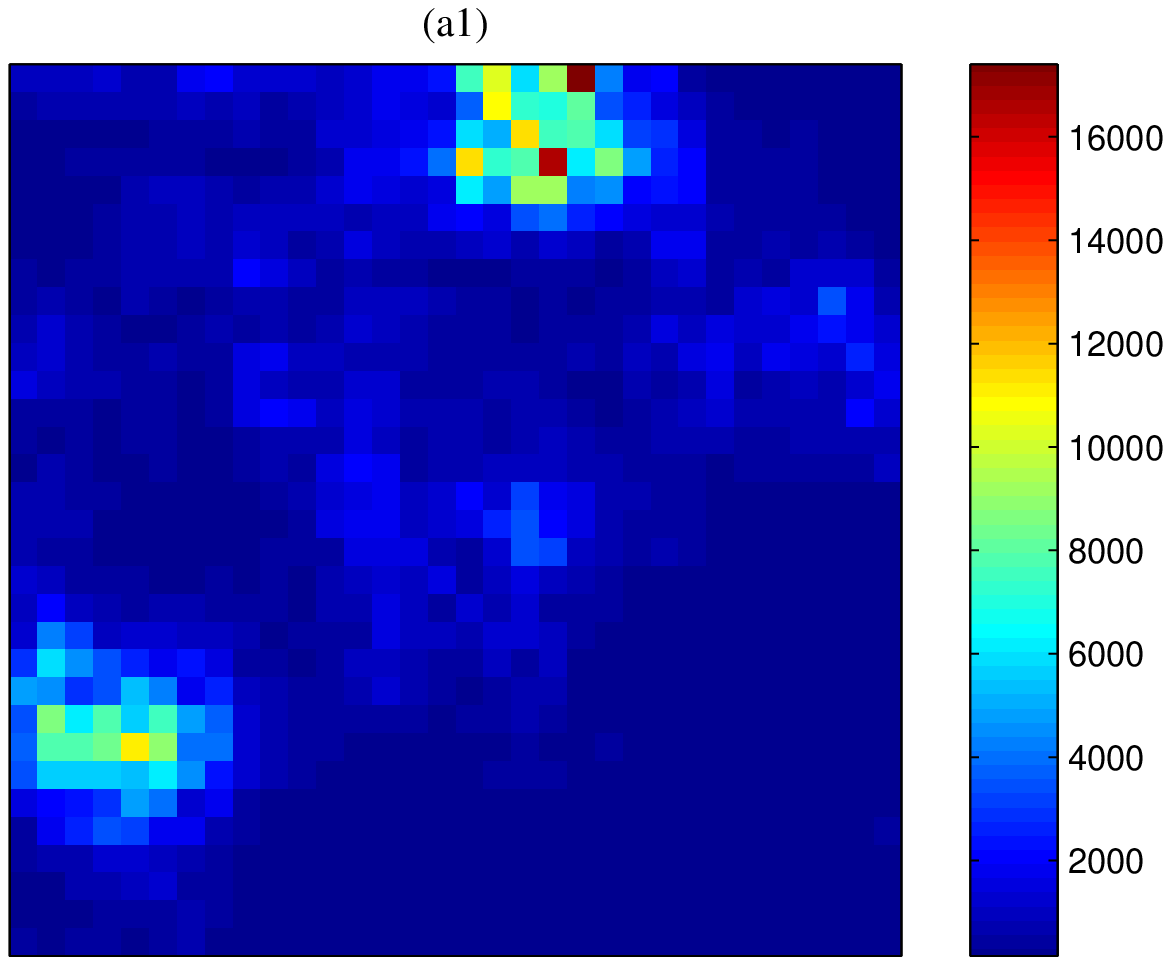}\includegraphics[%
  scale=0.25]{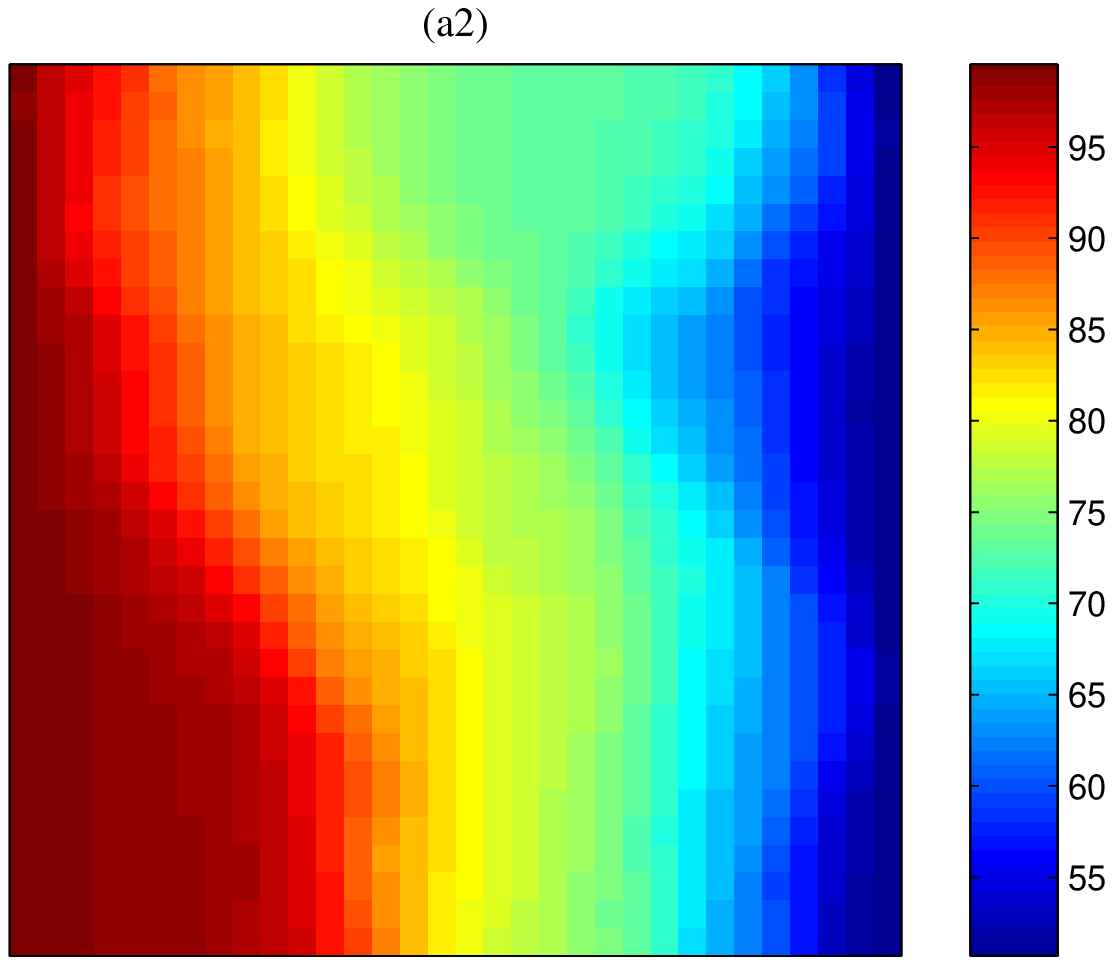}\includegraphics[%
  scale=0.25]{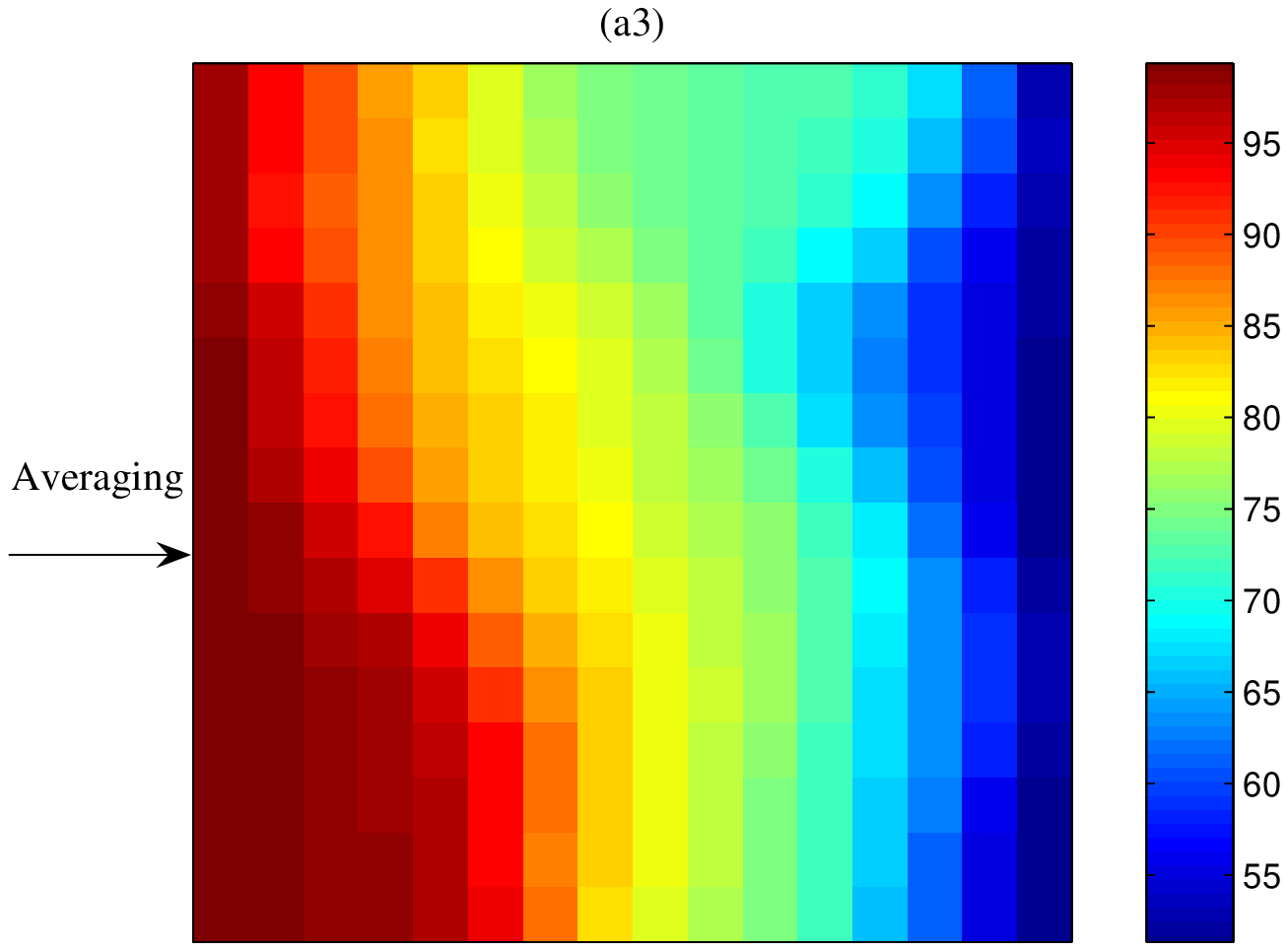}\end{center}

\begin{center}\includegraphics[%
  scale=0.25]{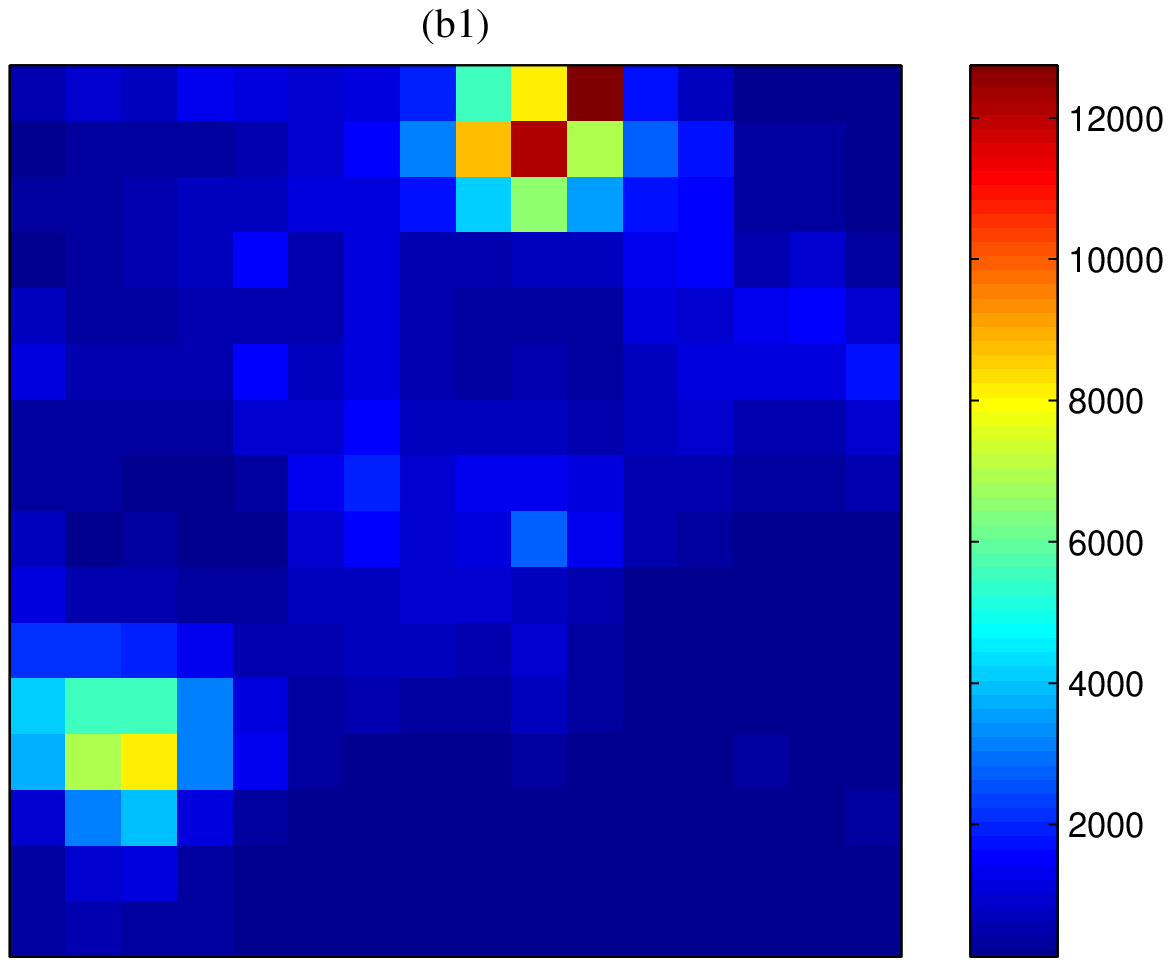}\includegraphics[%
  scale=0.25]{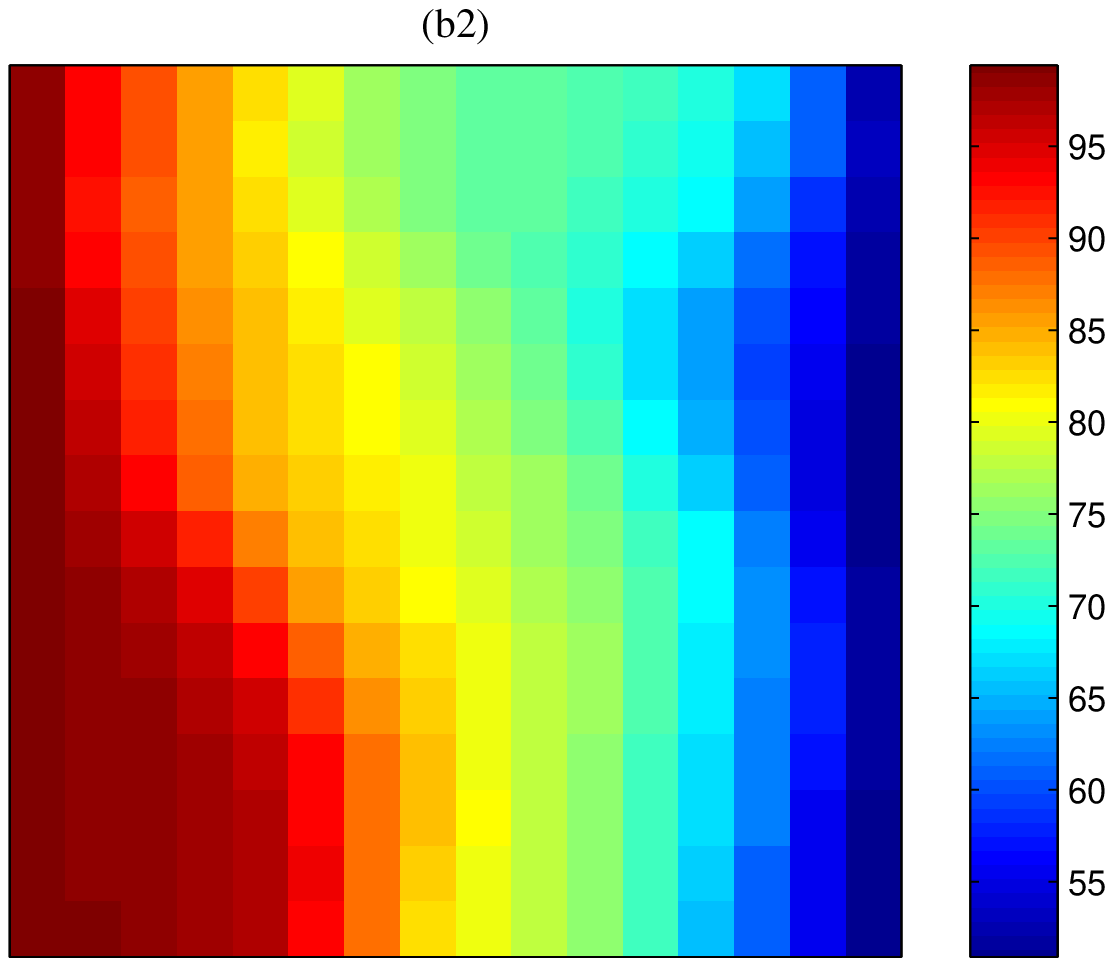}\includegraphics[%
  scale=0.25]{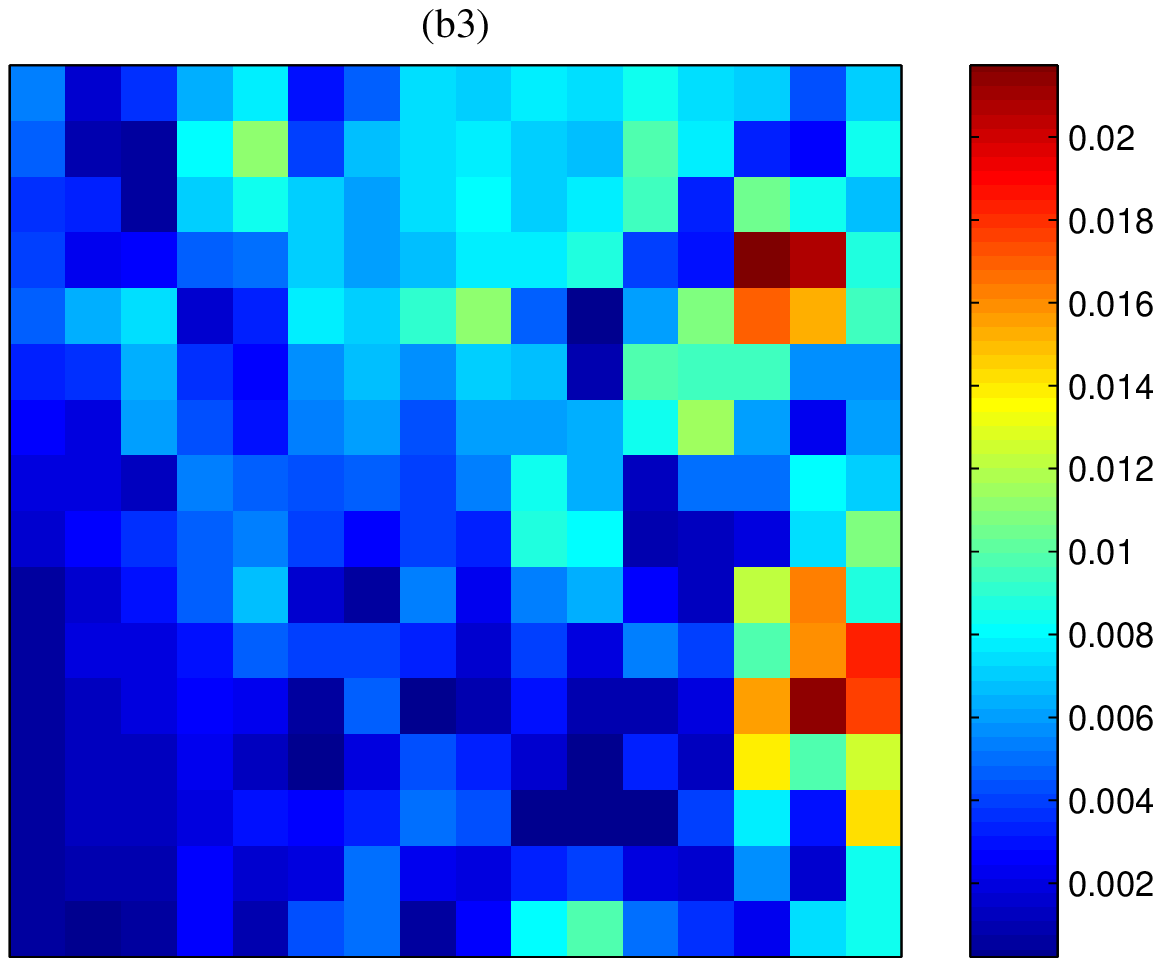}\end{center}

\caption{\label{cap:Wavelet-transform-based}Wavelet renormalization
of a permeability map from $32\times32$ to $16\times16$.
(\textbf{a1}) Fine scale permeability. (\textbf{a2}) Fine scale
pressure solution obtained from fine scale permeability.
(\textbf{a3}) Average of the fine scale pressure solution
($2\times2$ cells averaged). (\textbf{b1}) Wavelet renormalized
coarse permeability. (\textbf{b2}) Coarse pressure solution obtained
from coarse permeability. (\textbf{b3}) Modulus of relative error,
$\left|\textrm{a3-b2}\right|$/a3. In this case the relative
difference between the averaged fine scale pressure (\textbf{a3})
and the coarse pressure (\textbf{b2}) is within 2\%. This procedure
was repeated for systems with varying permeability ranges and with
different heterogeneities, simulating different rock types.}
\end{figure}

Once the permeability maps had been generated, pressure boundary
conditions were set on the left and right boundaries of the system.
These were taken to be fixed at 100 on the left and 50 on the right.
The drop in pressure across the system is a fundamental factor in
determining the errors in the estimates. However, the use of
relative errors mitigates this effect and the same boundary
conditions were used in all the simulations. A pressure profile was
obtained at each renormalization step inverting the corresponding
transmissibility matrix with the correct renormalized boundary
conditions and compared to an equally coarsened pressure obtained by
successively averaging fine pressure on $2\times2$ cells, see Figure
\ref{cap:Wavelet-transform-based}. The process was repeated $10$
times to generate a distribution of results.

When averaged over many realizations, the absolute error between the
averaged fine scale pressure and the coarse pressure obtained from
the wavelet upscale was consistently found to be of order $10^{-3}$.
For this kind of systems, errors in a single realizations did not
exceed 5\%.

As expected, the error was found to be higher with higher standard
deviation of the permeability, see Table
\ref{cap:Comparison-of-error r=3D1}, but only for very heterogeneous
systems, where the standard deviation is an order of magnitude
larger than the mean.

\begin{table}[H]
\begin{center}\begin{tabular}{ccc}
\hline $\sigma/\mu$& Mean relative error (10$^{-3}$)& Std of error
(10$^{-3}$)\tabularnewline \hline $\phantom{1}$0.1& $-$5.23&
3.41\tabularnewline $\phantom{1}$0.2& $-$0.74& 3.47\tabularnewline
$\phantom{1}$0.4& $-$0.84& 3.34\tabularnewline $\phantom{1}$0.8&
$-$1.46& 3.15\tabularnewline $\phantom{1}$1$\phantom{.1}$&
$\phantom{-}$0.58& 3.64\tabularnewline $\phantom{1}$2$\phantom{.1}$&
$\phantom{-}$1.82& 3.52\tabularnewline 10$\phantom{.1}$&
$\phantom{-}$0.79& 4.71\tabularnewline \hline
\end{tabular}\end{center}

\caption{\label{cap:Comparison-of-error r=3D1}Comparison of mean and
standard deviation of the relative error at different standard
deviation of permeability and same correlation length $r$=3,
$\mu$=10000, averaged over entire system. All data averaged over 27
realizations of $32\times32$ systems being upscaled to $16\times16$.
}
\end{table}

\begin{table}[H]
\begin{center}\begin{tabular}{ccc}
\hline Correlation length& Mean relative error (10$^{-3}$)& Std of
error (10$^{-3}$)\tabularnewline \hline 1& $-$0.52&
3.85\tabularnewline 2& $-$0.49& 3.51\tabularnewline 3&
$\phantom{-}$0.67& 3.46\tabularnewline 4& $\phantom{-}$0.64&
3.45\tabularnewline 5& $\phantom{-}$0.66& 3.43\tabularnewline \hline
\end{tabular}\end{center}

\caption{\label{cap:Comparison-of-error std=3D}Comparison of error
for different correlation lengths but same standard deviation,
$\sigma$=1000, $\mu=1000$ (average of multiple realizations of
$32\times32$ systems being upscaled to $16\times16$, see text for
details about the number of realizations). Notice a very weak
dependence of the standard deviation of the error on the correlation
length that seems to suggest that a more correlated system can be
upscaled more accurately.}
\end{table}

Next, a comparison between realizations with varying correlation
length $r$, expressed in terms of grid cells, and equal standard
deviation in permeability was made. A different number of
realizations were averaged depending on the correlation length of
the system, considering that each subsystem of linear size equal to
the correlation length constitutes a sample in statistical terms
(number of realizations = $3r^{2}$). As can be seen in Table
\ref{cap:Comparison-of-error std=3D}, the more the field is
correlated, that is, the larger the value of $r$, the better the
wavelet renormalization method approximates the fine scale pressure
average. However, even at a radius of correlation equal to one grid
cell, the average standard deviation of the error is within 0.4\%.

While the error averaged over the entire system can be misleadingly
small, due to cancellations which occur between positively and
negatively biased results at specific locations, the standard
deviation of the error over the system can be taken as a faithful
indicator of the performance of the method.

\begin{figure}
\begin{center}\includegraphics[%
  scale=0.25]{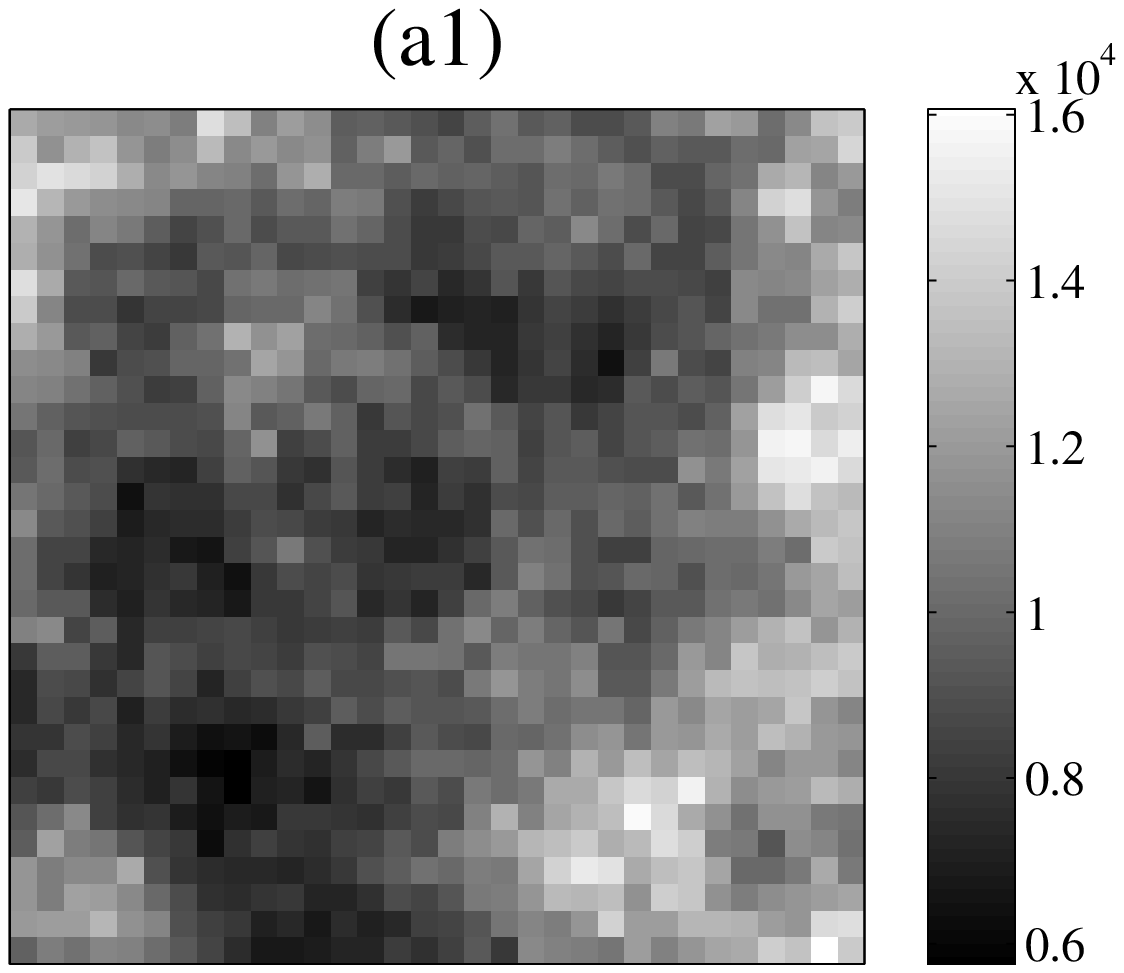}\includegraphics[%
  scale=0.25]{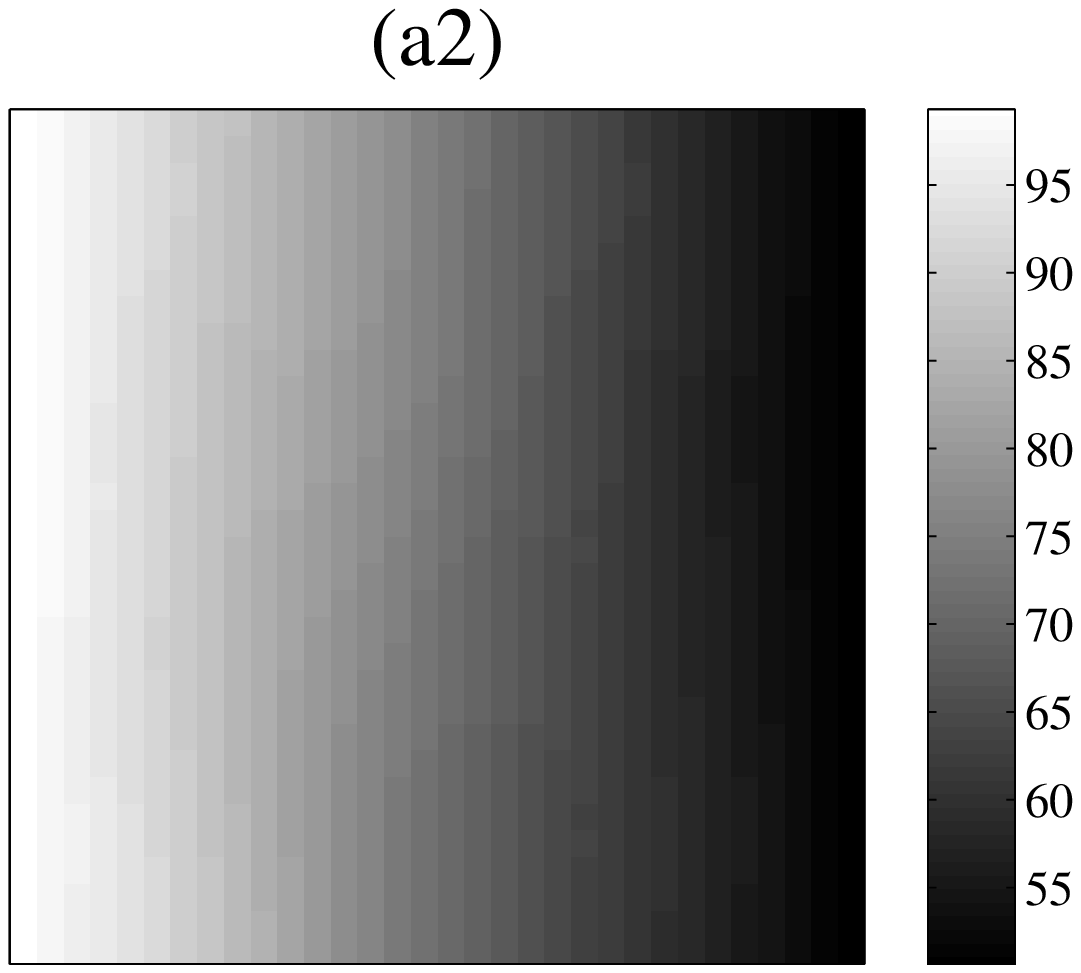}\end{center}

\begin{center}\includegraphics[%
  scale=0.25]{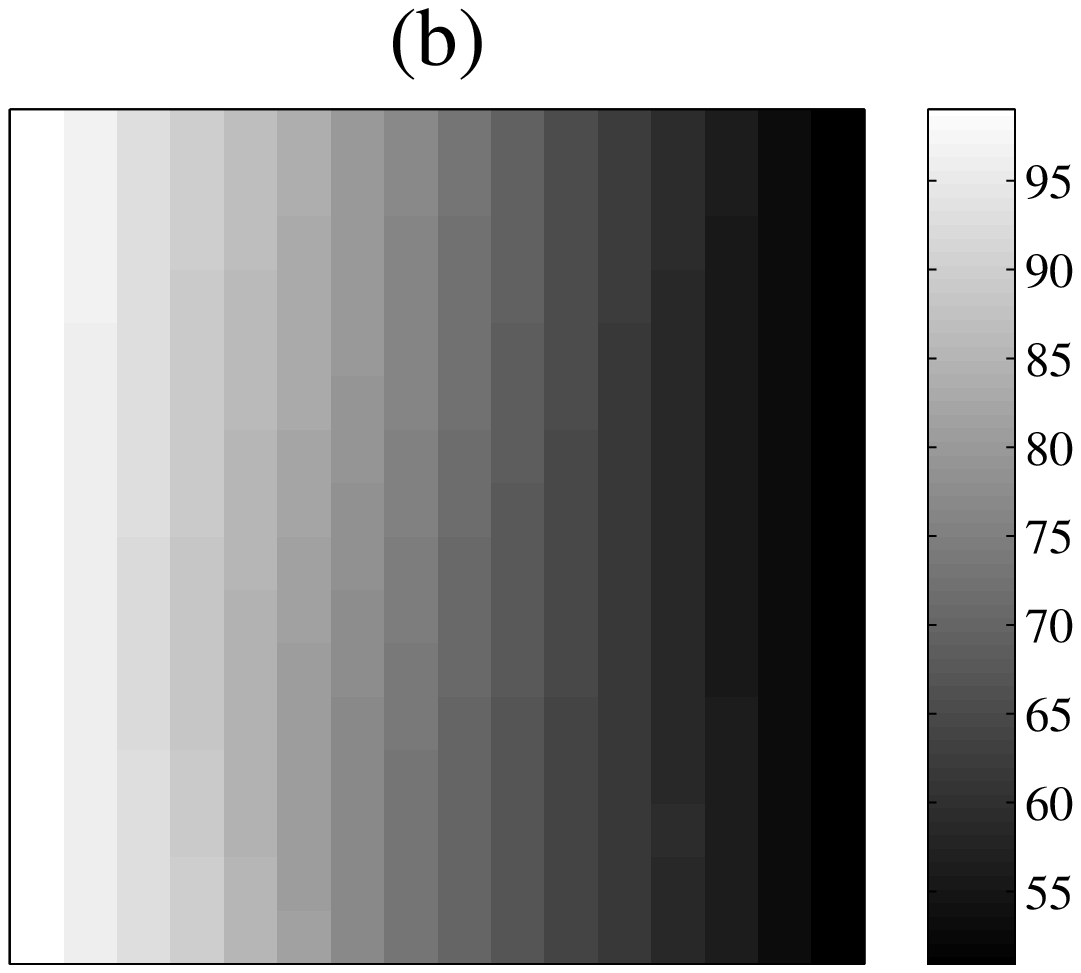}\includegraphics[%
  scale=0.25]{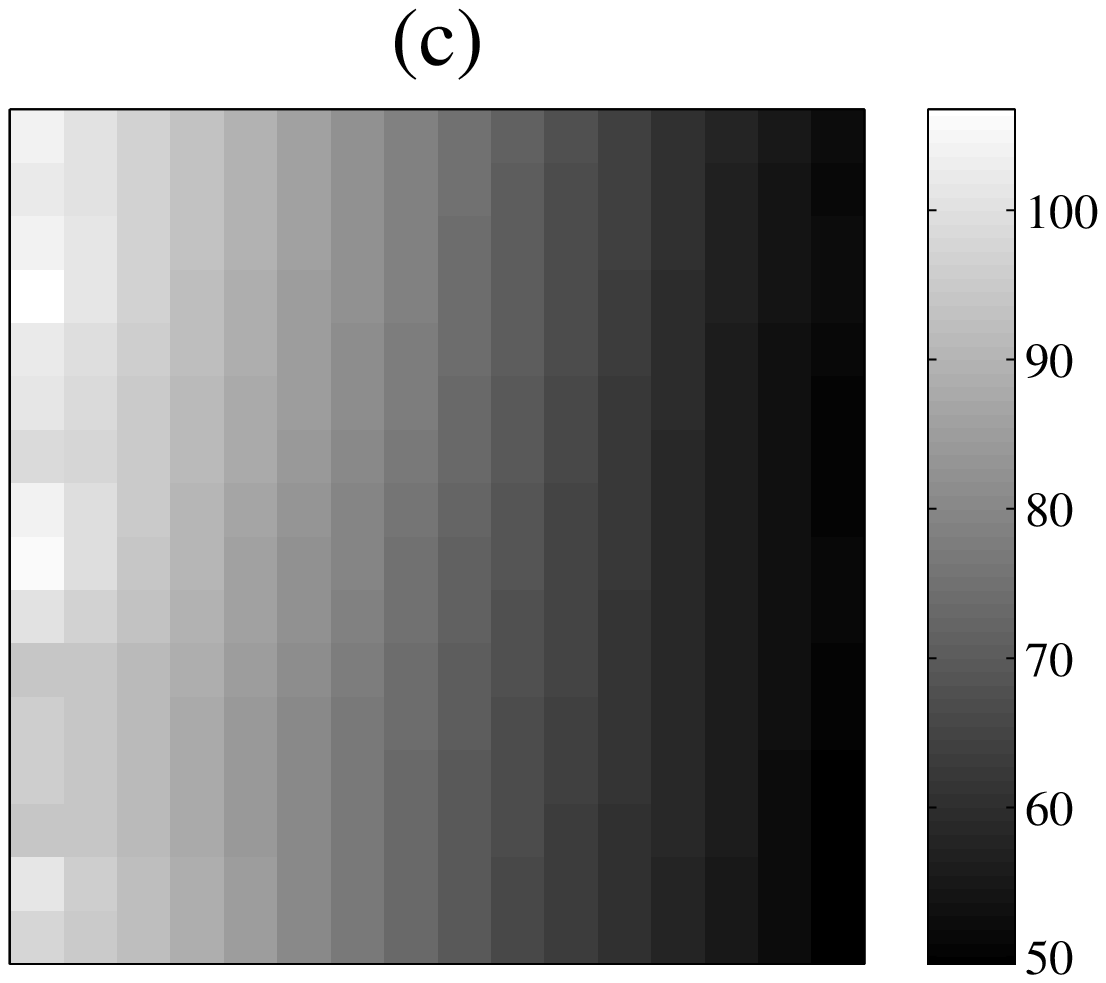}\end{center}

\begin{center}\includegraphics[%
  scale=0.25]{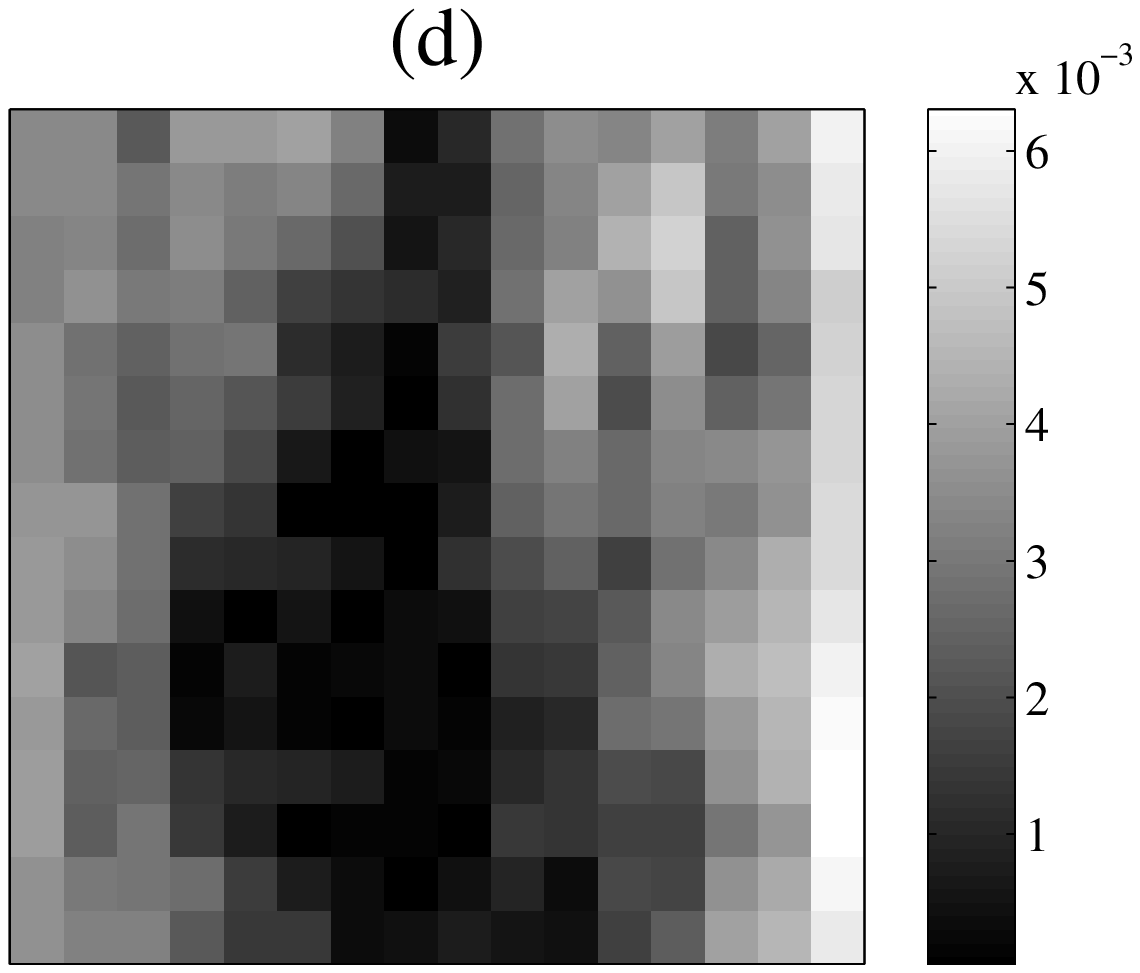}\includegraphics[%
  scale=0.25]{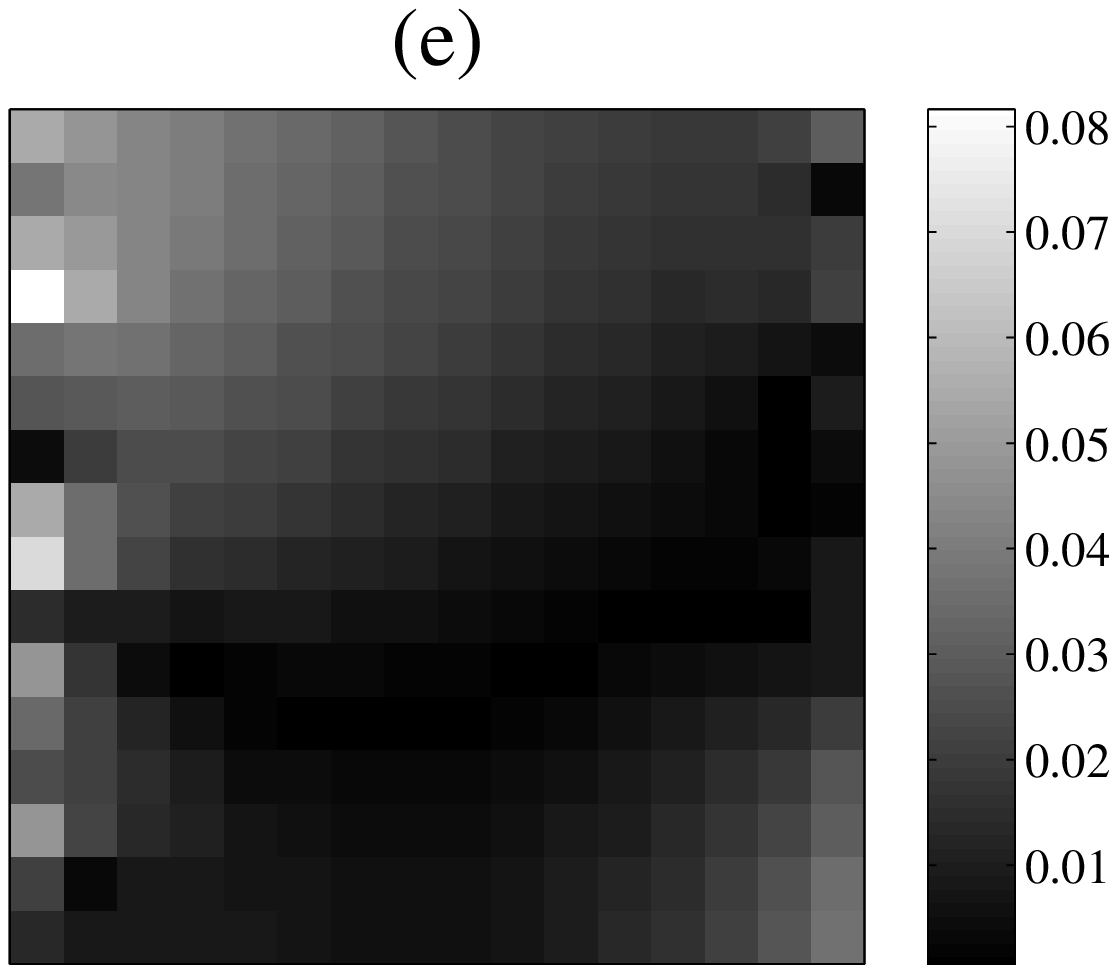}\end{center}

\caption{\label{cap:Wavres comp}Comparison of resistor and wavelet
renormalization. (\textbf{a}1) Fine scale permeability. (a2) Fine
scale pressure solution obtained from fine scale permeability. (b)
Coarse pressure solution obtained with wavelet method. (c) Coarse
pressure solution obtained with resistor method. (d) Modulus of
relative error of pressure obtained with the wavelet method with
respect to the pressure average. (e) Modulus of relative error of
pressure obtained with the resistor method with respect to the
pressure average. The error in the wavelet renormalization is of
order $10^{-3}$ because the permeability field is fairly
homogeneous. However, the resistor renormalization is less effective
even in this ideal case. }
\end{figure}

A comparison with the resistor renormalization performed according
to \cite{King:immis} (equation 2.1), can be seen in Figure
(\ref{cap:Wavres comp}). It is possible to develop a more accurate
resistor renormalization algorithm by considering the anisotropy
generated by the upscaling process. However, this algorithm is not
as immediate as the wavelet renormalization algorithm to implement,
requiring the definition of two transmissibilities per cell. It must
be noted that, while the wavelet method is geared towards
reproducing the average pressure, the resistor method is based on
flux conservation, thus it is not surprising that the results of the
two methods differ.

\subsection{Shales}

One of the major drawbacks of the renormalization proposed by
\cite{King:renkeff} is its imprecise treatment of shales. When the
permeability contrast between adjacent cells is high, for example at
the interface between permeable rock such as sandstone, and
impermeable elements such as shales, the analogy with resistors
gives inaccurate predictions. This results in a deformation of
shales which can lead to misjudgment of the reservoir connectivity.
Typically, shales have a large aspect ratio and they are distributed
horizontally often constituting a barrier to flow in the vertical
direction. A successful alternative approach to shales is given in
Ref. \cite{Begg:vertsh}, where the permeability is related to the
length of the path going around the shale bodies.

Shales were implemented in the following way: some of the sites of
the system were chosen at random and shales of random size and
aspect ratio were created by setting the permeabilities in the area
to a very small value ($10^{-13}$). Another conventional way of
implementing shales into a model is to make correlation very
anisotropic. This causes areas of low permeability to naturally
emerge with the correct aspect ratio and orientation. However, the
chosen method provides a much greater difference between the low
permeability of the shales and the distribution of the permeability
in the sand, which is often characteristic of physical systems.

\begin{figure}[H]
\begin{center}\includegraphics[%
  scale=0.25]{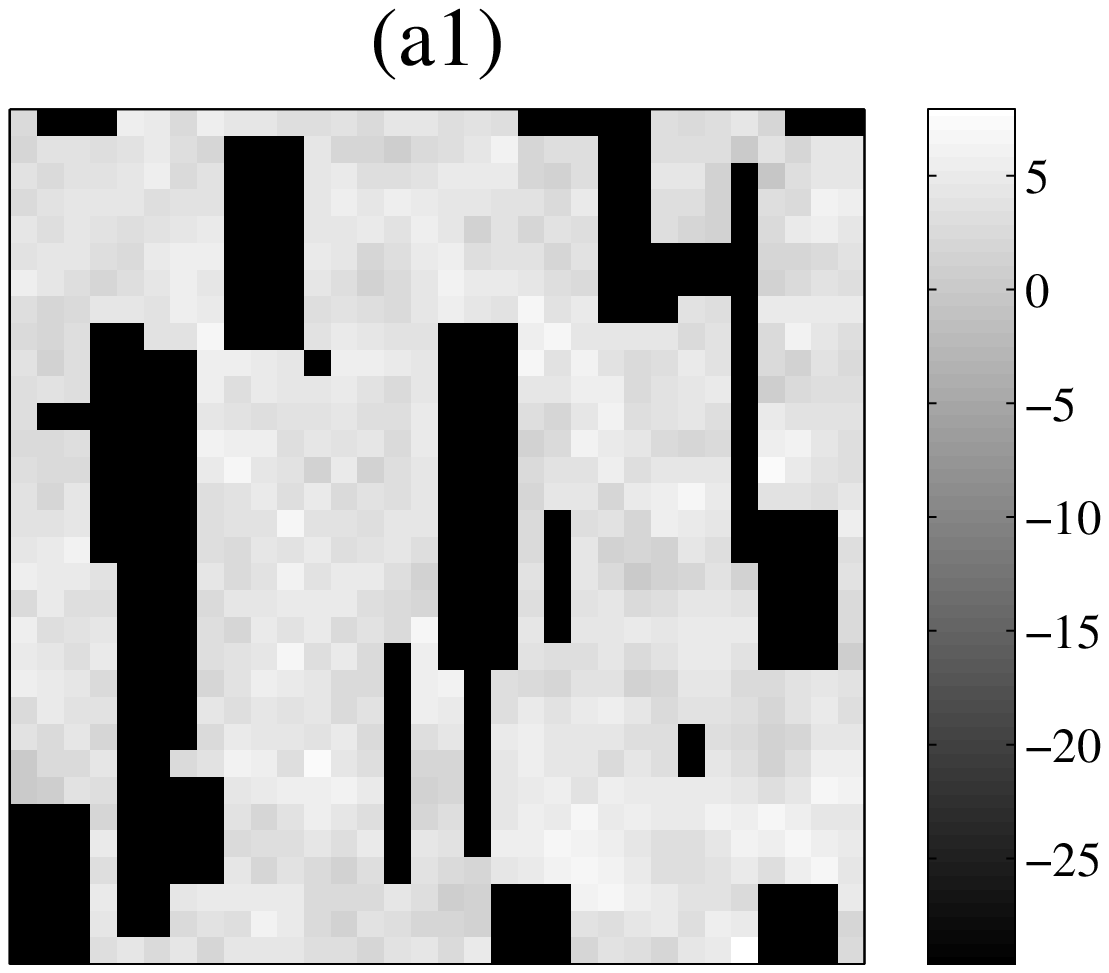}\includegraphics[%
  scale=0.25]{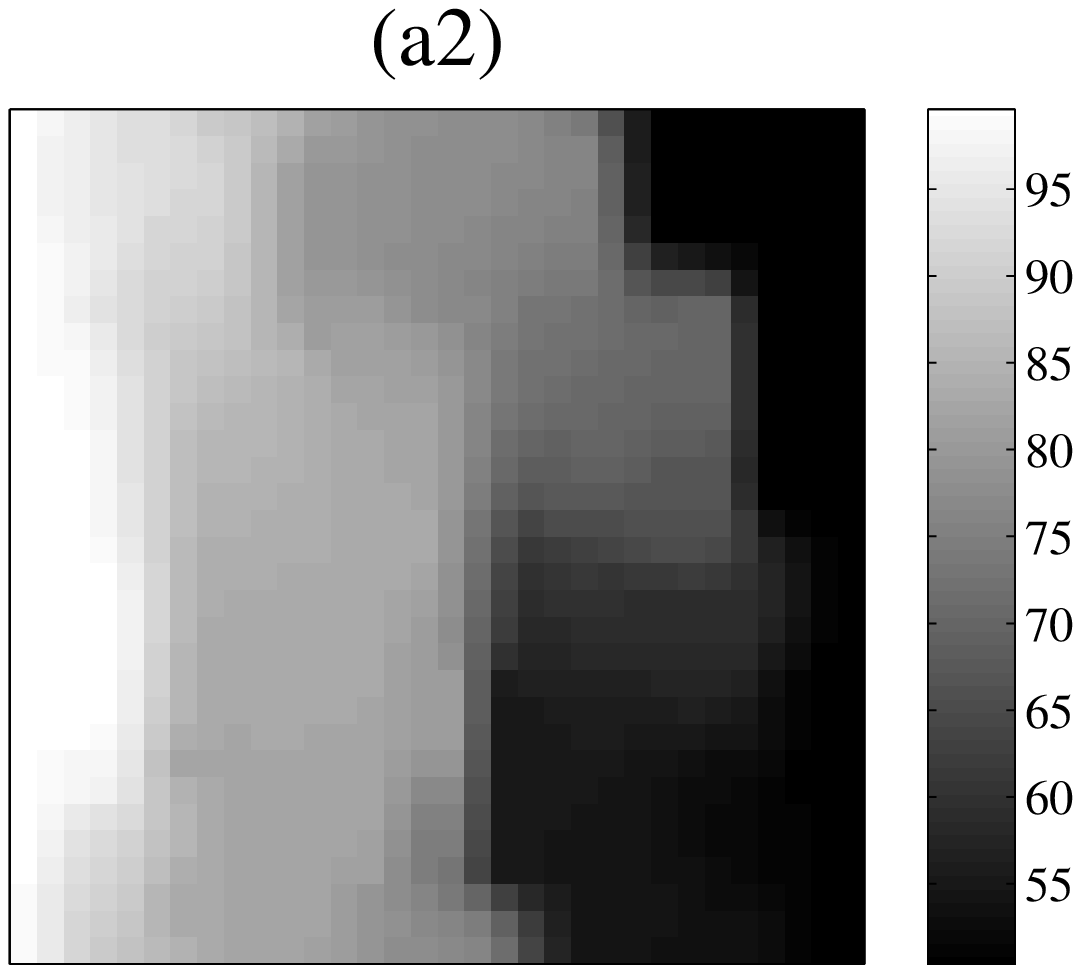}\includegraphics[%
  scale=0.25]{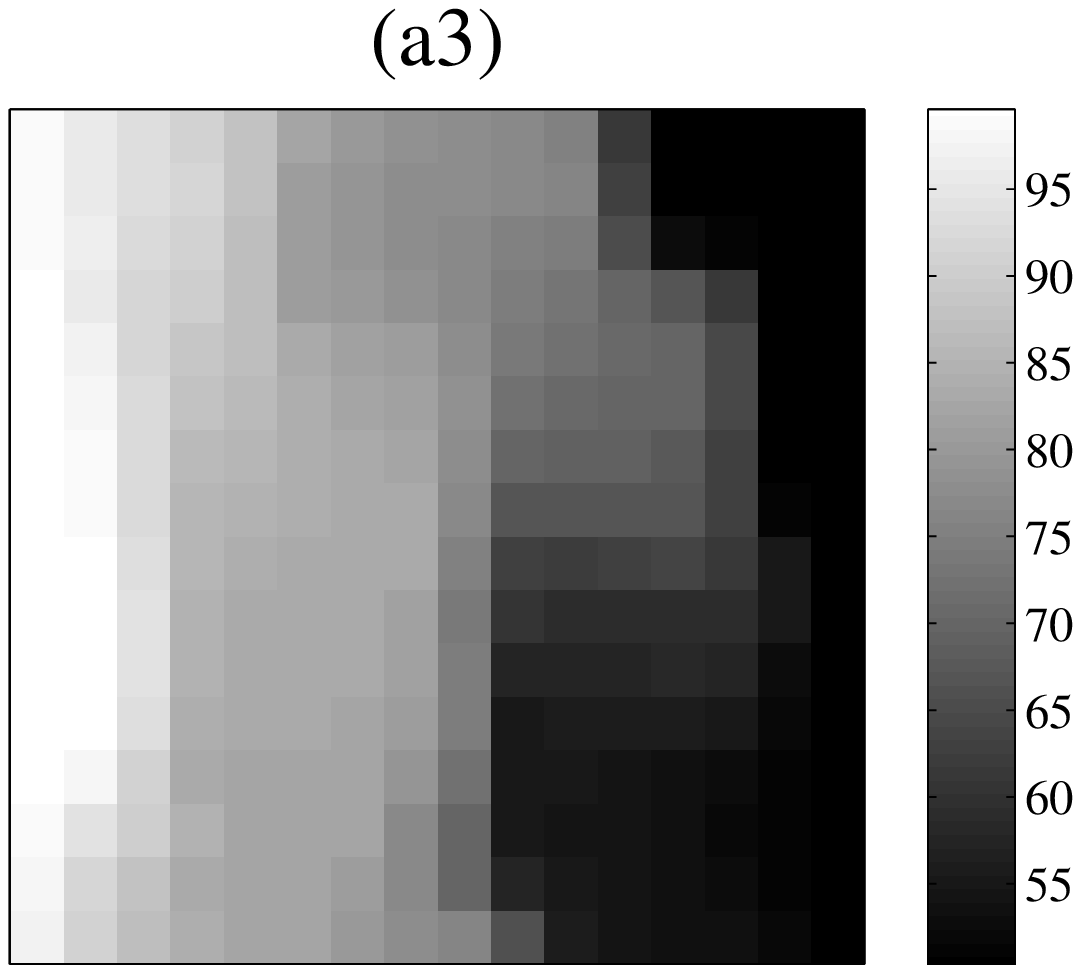}\end{center}

\begin{center}\includegraphics[%
  scale=0.25]{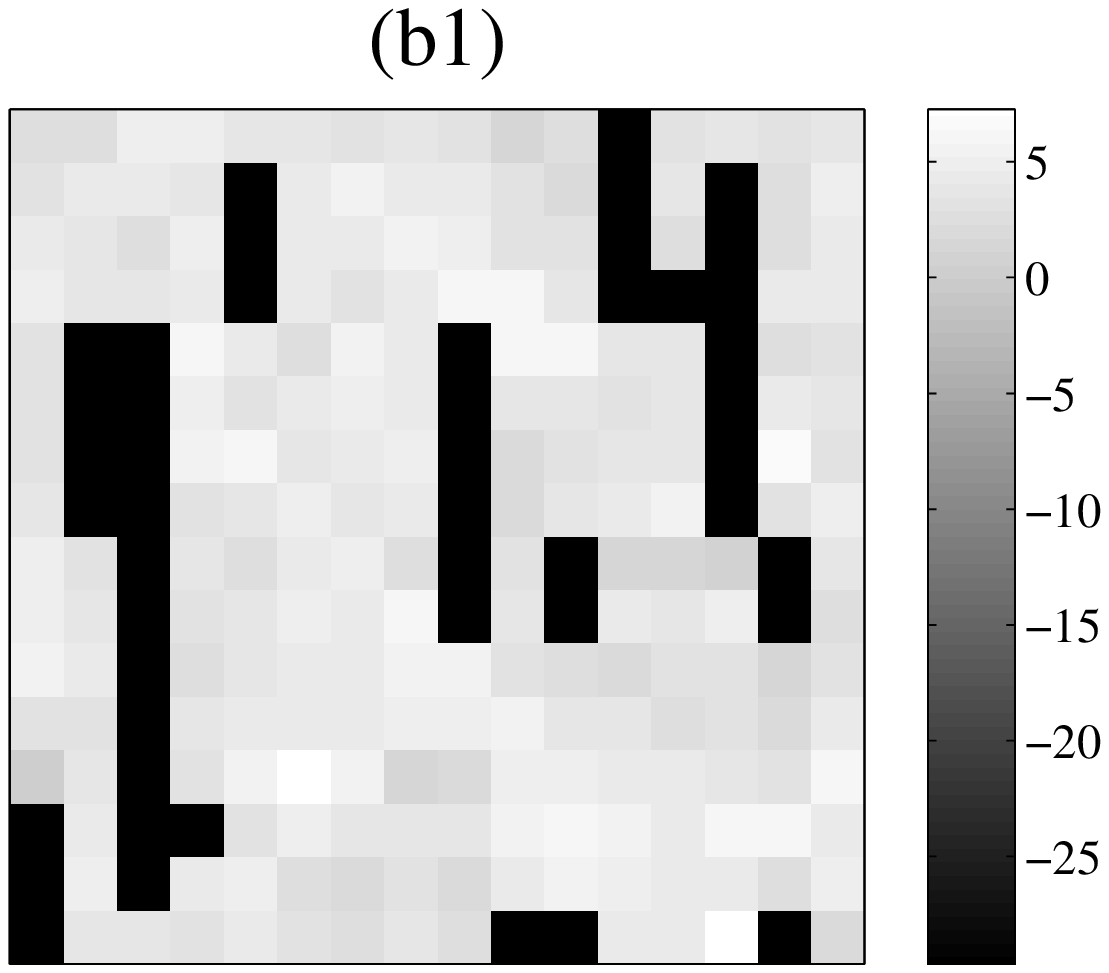}\includegraphics[%
  scale=0.25]{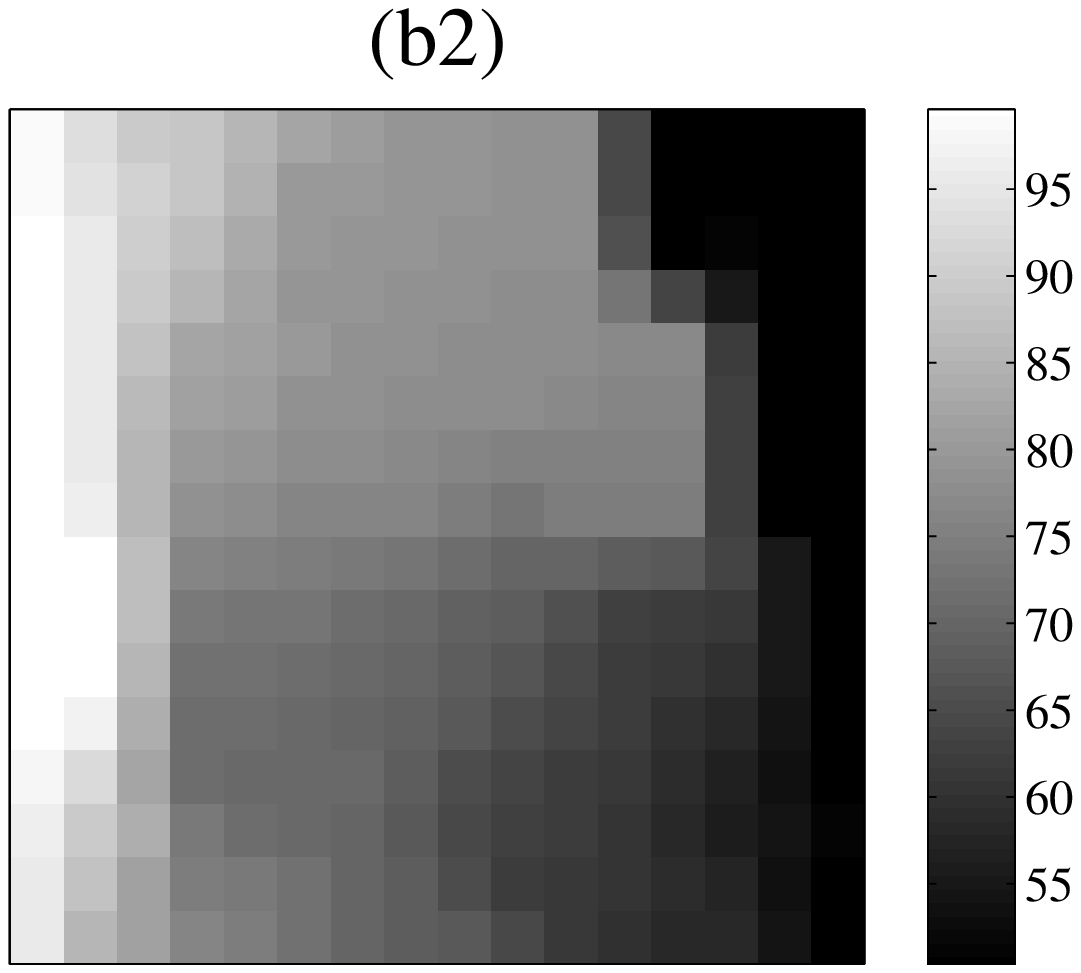}\includegraphics[%
  scale=0.25]{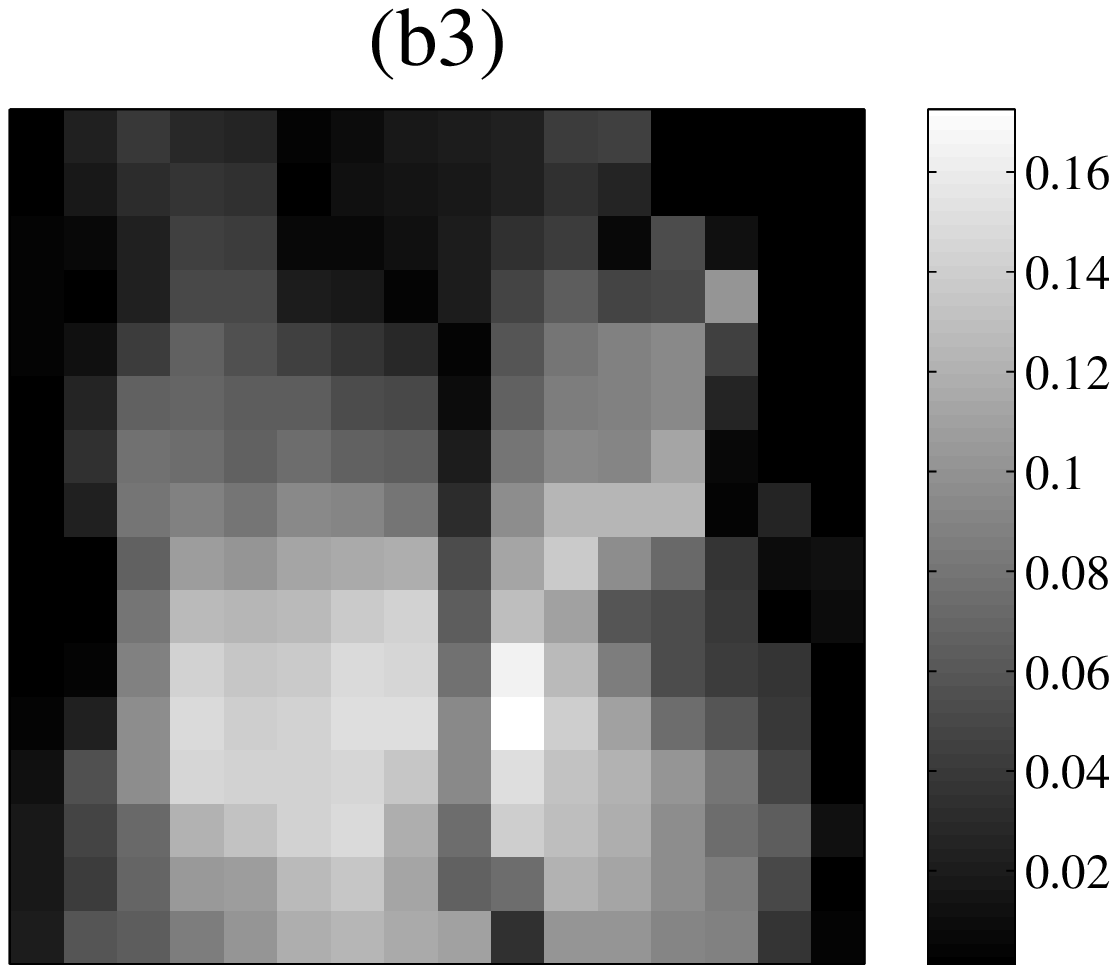}\end{center}

\begin{center}\includegraphics[%
  scale=0.25]{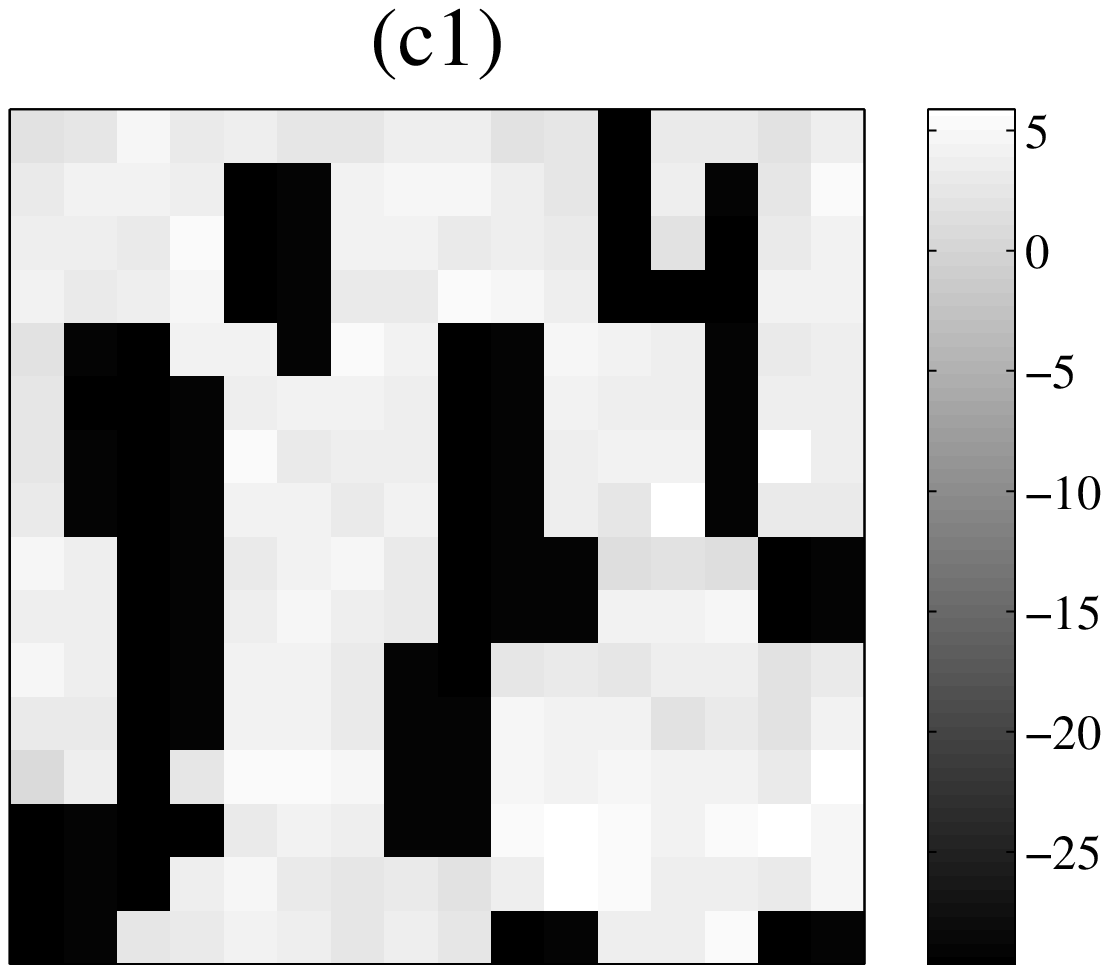}\includegraphics[%
  scale=0.25]{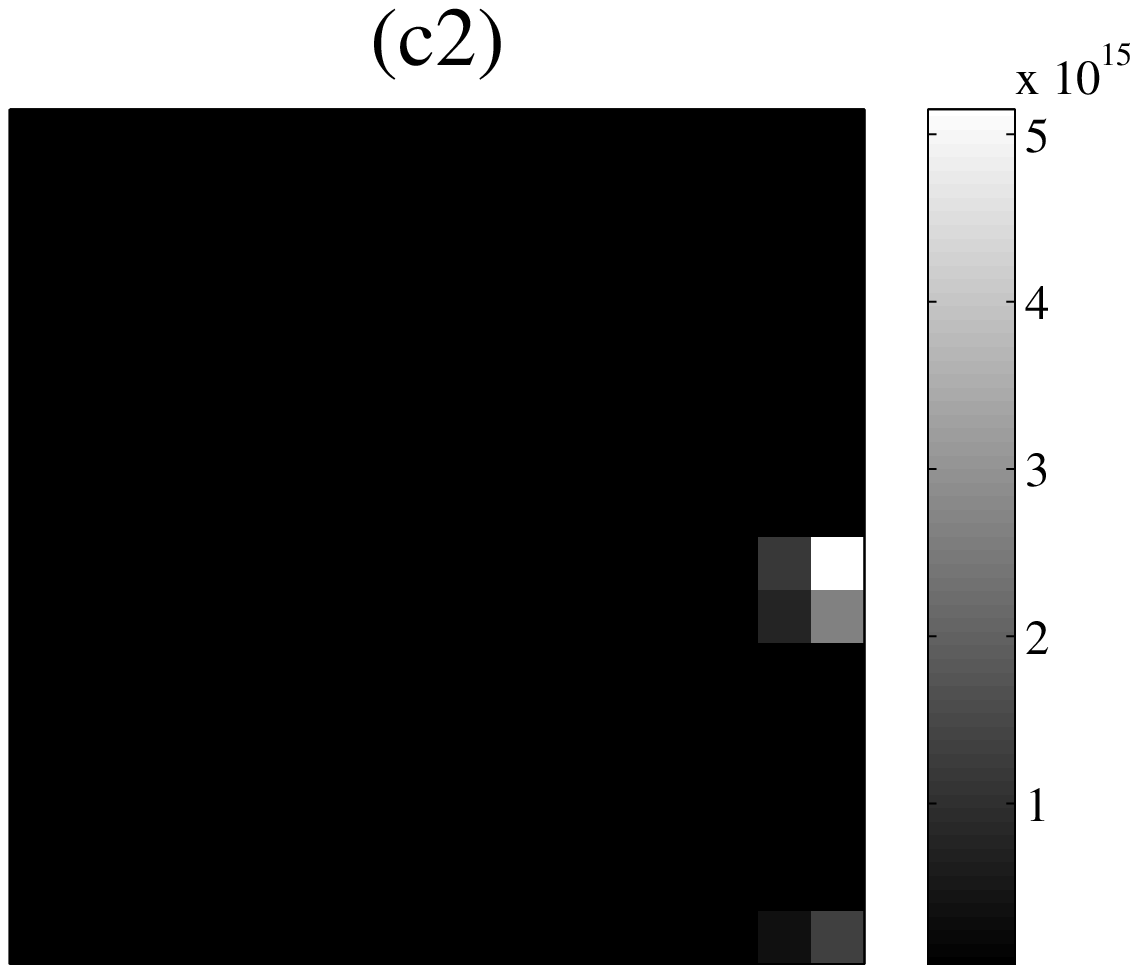}\includegraphics[%
  scale=0.25]{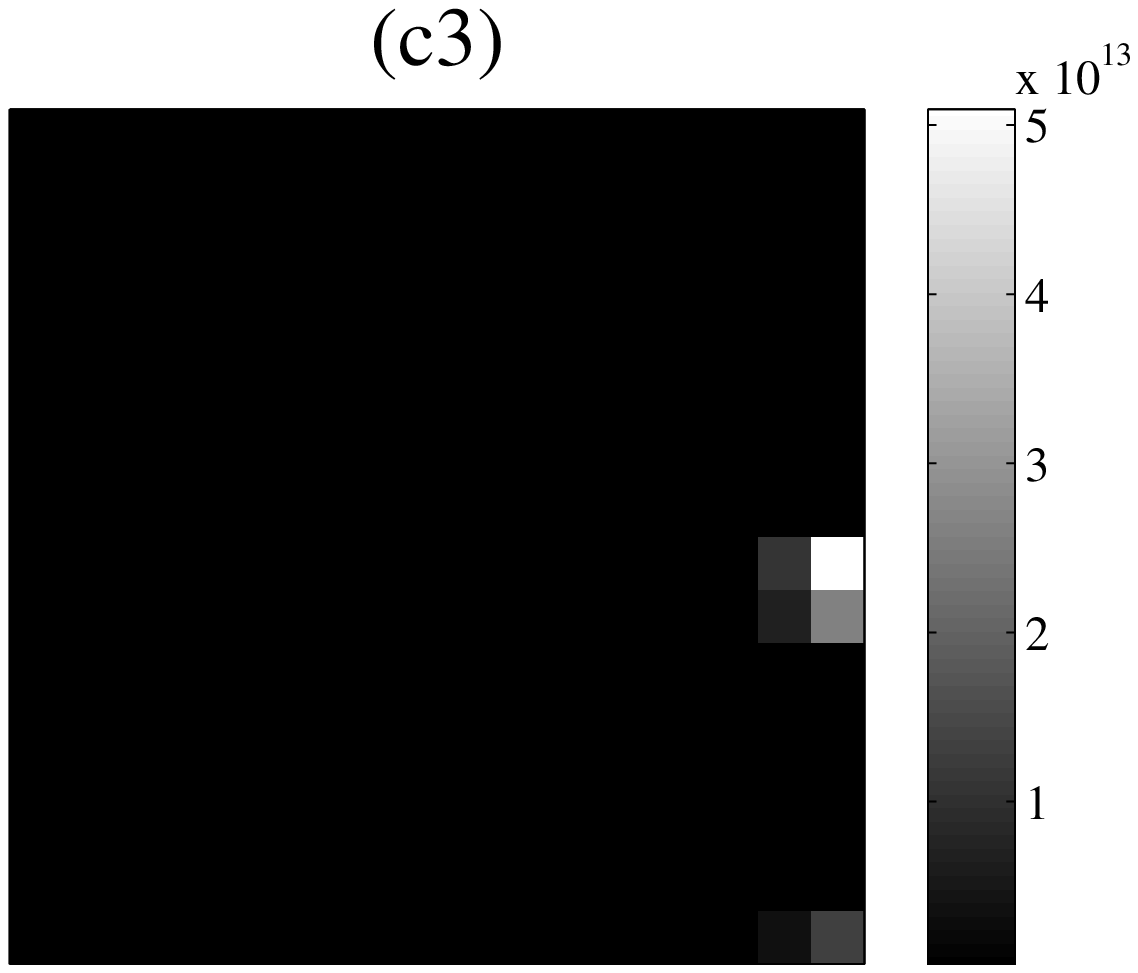}\end{center}

\caption{\label{cap:Shale error}Wavelet transform based real-space
renormalization of a permeability map with vertical shales from
$32\times32$ to $16\times16$. (\textbf{a1}) Fine scale permeability.
(\textbf{a2}) Fine scale pressure solution obtained from fine scale
permeability. (\textbf{a3}) Average of the fine scale pressure
solution ($2\times2$ cells averaged). (\textbf{b1})
Wavelet-renormalized coarse permeability. (\textbf{b2}) Coarse
pressure solution obtained from b1. (\textbf{b3}) Modulus of
relative error, $\left|\textrm{a3-b2}\right|$/a3. (\textbf{c1})
Resistor-renormalized coarse permeability. (\textbf{c2}) Coarse
pressure solution obtained from c1. (\textbf{c3}) Modulus of
relative error, $\left|\textrm{c3-c2}\right|$/c3. The relative
difference between the averaged fine scale pressure and the coarse
pressure from wavelet renormalization reaches 16\% with an average
of 6\%. Resistor renormalization clearly doesn't produce the
required result. The shale permeability is set to $10^{-13}$and also
the shales are distributed across the direction of flow, generating
a worst case scenario.}
\end{figure}

As can be seen in Figure \ref{cap:Shale error} and Table
\ref{cap:shalestable}, shales are correctly upscaled unless their
size becomes comparable to the size of the cells.

\begin{table}[H]
\begin{center}\begin{tabular}{ccccc}
\hline Max width& Max height& Shale fraction (\%)& Mean
error(10$^{-3}$)& Std of error(10$^{-3}$)\tabularnewline \hline
$\phantom{1}$2& $\phantom{1}$2& $\phantom{1}$3.2&
$\phantom{1}$18.29& $\phantom{1}$16.2\tabularnewline $\phantom{1}$2&
$\phantom{1}$2& 33.4& 113.85& 100.8\tabularnewline 16&
$\phantom{1}$5& 16.4& $\phantom{11}$9.1$\phantom{1}$&
$\phantom{1}$25.3\tabularnewline 16& $\phantom{1}$5& 33.4&
$\phantom{11}$6.73& $\phantom{1}$27.1\tabularnewline 16&
$\phantom{1}$5& 57.6& $\phantom{11}$3.1$\phantom{1}$&
$\phantom{1}$24.9\tabularnewline $\phantom{1}$5& 16& 18.8&
$\phantom{1}$48.7& $\phantom{1}$46$\phantom{1}$\tabularnewline
$\phantom{1}$5& 16& 36$\phantom{.1}$& $\phantom{1}$26.8&
$\phantom{1}$47.9\tabularnewline $\phantom{1}$5& 16& 52.2&
$\phantom{1}$20.3& $\phantom{1}$60$\phantom{1}$\tabularnewline
\hline
\end{tabular}\end{center}

\caption{\label{cap:shalestable}Error in up-scaling a system with
shales with different aspect ratio. Shale permeability set to
$10^{-13}$ . All values were averaged over 3 runs. Notice that
vertical and small shales are associated with a bigger error.}
\end{table}

When either the shale fraction or the sand fraction approaches the
percolation threshold, the error of the \emph{}wavelet method
calculated with respect to the average of the fine pressure solution
can be of order $10^{-1}$. In this case shales will either cover the
entire system or tend to disappear. Anisotropy also plays an
important role. Shales perpendicular to the flow seem to represent
more of a problem, since they oppose the pressure gradient, see
Table \ref{cap:shalestable}, bottom three entries. For example, in
Figure \ref{cap:Shale error}, the largest error occurs in the lower
central region where a vertical barrier disappears in the coarsening
process. However, in this situation, it is debatable that averaging
the pressure profile can be of any use. Visually, it is clear that
the upscaled pressure profile reproduces the fine scale pressure
profile with reasonable accuracy. The resistor renormalization, as
defined in Section \ref{sub:Analysis-of-heterogeneity}, produces
very unsatisfactory results. It must be noted that, even at the fine
scale, pressure in very nearly zero permeability areas is poorly
defined.

\subsection{Three-dimensional systems}

As already mentioned, the wavelet renormalization method for
up-scaling is easily extended to three-dimensional systems. In this
case, no flow was assumed in two directions and pressures were
specified on the boundaries of the third direction. The only
difference in the procedure between two- and three-dimensional
systems is the structure and size of the matrix $\mathbf{W}$. As for
the two-dimensional case, by observing the structure of the
transformed transmissibility matrix $\mathcal{T}$, a renormalization
scheme can be devised to produce the coarse permeability avoiding
the matrix multiplications. The algorithm substitutes cubes of
linear size 4 cells by cubes of half the linear size. Preliminary
runs confirm that the upscale procedure is approximately as accurate
as it is in the two-dimensional case.

\section{Conclusion}

An up-scaling method, based on the Haar wavelet transform and
real-space renormalization was presented. Its advantages are speed,
due to the underlying renormalization algorithm, and a rigorous
mathematical derivation of the up-scaling rule.

This algorithm emerges from a mean-field picture of the solution to
Darcy's equation, which is at the heart of the success of
renormalization methods. The renormalization scheme is a consequence
of the choice of $\mathbf{W}$ matrix, in this case the aim is to
obtain the coarse permeability map that would generate the average
pressure profile. A different matrix would lead to mathematically
valid results, for example one where the block permeability is taken
to be equal to the value at the top left cell in the group
constituting the block. However, the present choice attempts to
minimize the information loss inherent in the coarsening process
while preserving the algorithm simplicity to ensure its efficiency.

Within this context, the lowest degree, mean-field approximation, in
which all fluctuations are neglected, performs well in two and three
dimensions. The main problems with this method are encountered when
there is a high contrast in permeability, such as in the case of
shales, which leads to sharp pressure changes that inevitably get
smoothened out. The resistor renormalization fails even more
drastically in this case. A different wavelet matrix choice would
improve the performance of the method. It is nevertheless
foreseeable that the emerging renormalization scheme would not be as
easy to implement as the one presented. An exact solution could also
be obtained, including all the fluctuation terms, however, the
computational power required would be equivalent to performing the
fine-scale solution.

At present, the method can be used as a fast upscaling technique
able to cope with heterogeneities. The formalism introduced
highlights how a very crude renormalization scheme is satisfactory
in treating sufficiently homogeneous systems and how upscaling
methods can be constructed to match the specifications of the
problem and the required results. The resistor method is based on an
analogy with current laws and is therefore a statement of
conservation of flux. It is possible that by defining Darcy's
equation in terms of fluxes rather than permeability and pressure,
one might be able to find a matrix analogy to the resistor method
that will reproduce the resistor upscaling rule in the same way as
the current $\mathbf{W}$ matrix produces the renormalization scheme
that was proposed. The present framework can be applied to other
problems, such as advective transport, leading to insights into the
general issue of how operators change as a consequence of
coarsening.

It is hoped that further study will shed light on the effect of
adding fluctuations to the mean-field approximation, allowing the
choice between different degrees of accuracy depending on the
available computational time.

\section{Acknowledgements}

V.P. gratefully acknowledges funding from the Department of Earth
Science and Engineering, Imperial College London and the authors are
thankful to the anonymous referees for very useful comments.

\begin{landscape}
\section{Appendix}

In the following we reproduce the structure of the matrices
discussed in the text. The structure of the transmissibility matrix
for a $4\times4$ system:

\begin{flushleft}$\mathbf{T}=\left[\begin{array}{cccccccc}
2k_{1}+t_{1,2}+t_{1,5} & -t_{1,2} & 0 & 0 & -t_{1,5} & 0 & ... & 0\\
-t_{2,1} & 2k_{2}+t_{2,3}+t_{2,5} & -t_{2,3} & 0 & 0 & -t_{2,5} & ... & 0\\
... & ... & ... & ... & ... & ... & ... & -t_{15,16}\\
0 & 0 & 0 & 0 & 0 & . & -t_{16,15} &
2k_{16}+t_{16,15}+t_{16,12}\end{array}\right]$\end{flushleft}

\noindent \begin{flushleft}The upper corner of the transformed
matrix: $\mathbf{\mathcal{T}}=\mathbf{WTW^{T}}$\end{flushleft}

\noindent \begin{flushleft}\[
\mathbf{\mathcal{T}}=\left[\begin{array}{cccc}
k_{1}+k_{2}+\frac{t_{23}+t_{67}}{2}+\frac{t_{59}+t_{610}}{2} & -\frac{t_{23}+t_{67}}{2} & -\frac{t_{59}+t_{610}}{2} & 0\\
-\frac{t_{23}+t_{67}}{2} & k_{3}+k_{4}+\frac{t_{23}+t_{67}}{2}+\frac{t_{711}+t_{812}}{2} & 0 & -\frac{t_{711}+t_{812}}{2}\\
-\frac{t_{59}+t_{610}}{2} & 0 & k_{13}+k_{14}+\frac{t_{59}+t_{610}}{2}+\frac{t_{1011}+t_{1415}}{2} & -\frac{t_{1011}+t_{1415}}{2}\\
0 & -\frac{t_{711}+t_{812}}{2} & -\frac{t_{1011}+t_{1415}}{2} &
k_{15}+k_{16}+\frac{t_{711}+t_{812}}{2}+\frac{t_{1011}+t_{1415}}{2}\end{array}\right]\]
\end{flushleft}

The transmissibility matrix for a $2\times2$ system, the dash
indicates that the properties refer to the $2\times2$ system :

\noindent \begin{flushleft}\[ \mathbf{T'}=\left[\begin{array}{cccc}
2k'_{1}+t'_{1,2}+t'_{1,3} & -t'_{1,2} & -t'_{1,3} & 0\\
-t'_{2,1} & 2t'_{2}+t'_{2,1}+t'_{2,4} & 0 & -t'_{2,4}\\
-t'_{1,3} & 0 & 2t'_{3}+t'_{3,1}+t'_{3,4} & -t'_{3,4}\\
0 & -t'_{2,4} & -t'_{3,4} &
2k'_{4}+t'_{2,4}+t'_{3,4}\end{array}\right]\]
\end{flushleft}

Relationship between permeability and transmissibility in the
upscaled system ($k'_{i}$, $t'_{ij}$) and in the fine scale system
($k{}_{i}$, $t{}_{ij}$):

\[
k'_{1}=\frac{k_{1}+k_{2}}{2},\,\,\,\,
t'_{12}=\frac{t_{23}+t_{67}}{2},\,\,\,\,
t'_{13}=\frac{t_{59}+t_{610}}{2}\,\,\,\,\textrm{etc}.\]

\end{landscape}


\bibliographystyle{amsplain}
\bibliography{biblio}

\end{document}